\documentclass[twocolumn,aps,superscriptaddress]{revtex4-1}
\usepackage[dvips]{graphicx}
\usepackage{latexsym,amssymb,amsmath}
\usepackage{color}
\usepackage{enumerate}
\usepackage[bookmarksnumbered,bookmarksopen,colorlinks,citecolor=red,linkcolor=blue]{hyperref}
\usepackage{mathrsfs}
\usepackage{times}
\usepackage{bm}
\usepackage{csquotes}
\usepackage{multirow}
\usepackage{diagbox}
\usepackage{booktabs}
\usepackage{here}
\usepackage[dvipsnames]{xcolor}
\usepackage[normalem]{ulem}

\graphicspath{{./figs/}}
\renewcommand\sout{\bgroup \color{red} \ULdepth=-.5ex \ULset}

\begin{document}   

\title{Hadronization of Heavy Quarks}
\author{Jiaxing Zhao$^{1}$, J\"org  Aichelin$^{1}$, Pol Bernard Gossiaux}
\affiliation{SUBATECH, Nantes University, IMT Atlantique, IN2P3/CNRS, 4 rue Alfred Kastler, 44307 Nantes cedex 3, France}
\author{Andrea Beraudo}
\affiliation{Istituto Nazionale di Fisica Nucleare - Sezione di Torino, Via Pietro Giuria 1, I-10125, Torino, Italy}
\author{Shanshan Cao}
\affiliation{Institute of Frontier and Interdisciplinary Science, Shandong University, Qingdao, Shandong, 266237, China}
\author{Wenkai Fan}
\affiliation{Department of Physics, Duke University, Durham, North Carolina 27708, USA}
\author{Min He}
\affiliation{Department of Applied Physics, Nanjing University of Science and Technology, Nanjing 210094, China}
\author{Vincenzo Minissale}
\affiliation{Università degli Studi di Catania - Dipartimento di Fisica e Astronomia "Ettore Majorana", Catania, Italy}
\affiliation{INFN-Laboratori Nazionali del Sud, Catania, Italy}
\author{Taesoo Song}
\affiliation{GSI Helmholtzzentrum f\"{u}r Schwerionenforschung GmbH, Planckstrasse 1, 64291 Darmstadt, Germany}
\author{Ivan Vitev}
\affiliation{Theoretical Division, Los Alamos National Laboratory, Los Alamos, NM, 87545, USA}
\author{Ralf Rapp}
\affiliation{Cyclotron Institute and Department of Physics and Astronomy, Texas A\&M University, College Station, Texas 77843-3366, USA}
\author{Steffen Bass}
\affiliation{Department of Physics, Duke University, Durham, North Carolina 27708, USA}
\author{Elena Bratkovskaya}
\affiliation{GSI Helmholtzzentrum f\"{u}r Schwerionenforschung GmbH, Planckstrasse 1, 64291 Darmstadt, Germany}
\affiliation{Institute for Theoretical Physics, Johann Wolfgang Goethe Universit\"{a}t, Frankfurt am Main, Germany}
\affiliation{Helmholtz Research Academy Hessen for FAIR (HFHF),GSI Helmholtz Center for Heavy Ion Research. Campus Frankfurt, 60438 Frankfurt, Germany}
\author{Vincenzo Greco}
\affiliation{Università degli Studi di Catania - Dipartimento di Fisica e Astronomia "Ettore Majorana", Catania, Italy}
\affiliation{INFN-Laboratori Nazionali del Sud, Catania, Italy}
\author{Salvatore Plumari}
\affiliation{Università degli Studi di Catania - Dipartimento di Fisica e Astronomia "Ettore Majorana", Catania, Italy}
\affiliation{INFN-Laboratori Nazionali del Sud, Catania, Italy}
\date{\today}

\begin{abstract}
Heavy-flavor hadrons produced in ultra-relativistic heavy-ion collisions are a sensitive probe for studying hadronization mechanisms of the quark-gluon-plasma.
In this work, we survey how different transport models for the simulation of heavy-quark diffusion through a quark-gluon plasma in heavy-ion collisions implement hadronization and how this affects final-state observables. Utilizing the same input charm-quark distribution in all models at the hadronization transition, we find that the transverse-momentum dependence of the nuclear modification factor of various charm hadron species has significant sensitivity to the hadronization scheme. In addition, the charm-hadron elliptic flow  
exhibits a nontrivial dependence on the elliptic flow of the hadronizing partonic medium. 
\end{abstract}
\maketitle 

%==================================
\section{Introduction}
%=================================
At high temperatures and vanishing net-baryon density, lattice-discretized simulations of Quantum Chromodynamics (QCD) at finite temperature~\cite{Bernard:2004je,Cheng:2006qk} predict a transition of hadronic matter to a new phase of matter, called quark-gluon plasma (QGP), which is a state of deconfined quarks and gluons. This transition has been realized experimentally via heavy-ion collisions at the Relativistic Heavy-Ion Collider (RHIC) and the Large Hadron Collider (LHC). 
It is widely accepted by now that, upon initial impact of the incoming nuclei,  the system briefly evolves through a pre-equilibrium stage before a locally thermalized QGP is formed. Large pressure gradients in the fireball drive the QGP to expand and cool rapidly. Once the system reaches the crossover region with temperatures around the pseudo-critical temperature, $T_{\rm pc}$, the partons constituting the QGP undergo color-neutralization into hadronic bound states -- a process generically denoted as hadronization.

Quantitative studies of the properties of hot QCD matter are a central objective of ultrarelativistic heavy-ion collisions, which, however, require suitable probes that can be observed in the final state while still carrying information about the hot and dense phase phases. 
A promising probe in this regard is heavy-flavor (HF) hadrons
which have several advantageous features~\cite{Rapp:2009my}:
\begin{itemize}
\item  The heavy-quark (HQ) mass is much larger than the nonperturbative QCD scale, $m_c,m_b\gg \Lambda_{QCD}$, and thus the production in nuclear collisions mainly occurs in initial hard scatterings, which can be reasonably well described by perturbative QCD; 
\item The HQ mass is much larger than the typical temperature of the hot medium; this implies that their number is expected to be effectively conserved during the evolution due to high production thresholds and that their diffusion is akin to a Brownian motion which enables rather direct access to pertinent transport parameters.   
\item The determination of HQ fragmentation functions can be carried out at next-to-leading order in the production process within an HQ mass expansion using the methods of HQ effective theory (HQET)~\cite{Braaten:1994bz}.
\end{itemize}

Several research groups have developed transport approaches that describe the dynamics of heavy quarks from their creation at the onset of a heavy-ion collision through their evolution in the QCD medium until their detection as heavy hadrons (or their decay products) in experiment~\cite{vanHees:2005wb,He:2011qa,Minissale:2020bif,Cao:2015hia,Cao:2016gvr,Cao:2018ews,Cao:2019iqs,Gossiaux:2009mk,Song:2015sfa,Song:2015ykw,He:2019vgs,Li:2020zbk,Beraudo:2022dpz}. After about two decades of developments, these models are largely able to describe key experimental data, such as the scaled ratio between the transverse momentum ($p_T$) spectra in heavy-ion collisions and proton-proton collisions, $R_{AA}(p_T)$, and the azimuthal asymmetry in the $p_T$ spectra quantified by the elliptic-flow coefficient, $v_2(p_T)$, which have become increasingly precise in recent years~\cite{STAR:2017kkh,STAR:2018zdy,ALICE:2021rxa,ALICE:2020iug,CMS:2020bnz,CMS:2017qjw}.
These transport approaches have become rather complex, aimed at a comprehensive description of the initial production of the heavy quarks (including nuclear effects in the initial state), their interactions with the QGP, hadronization, and the interactions of heavy hadrons in the hadronic phase. Most of these processes cannot be rigorously described using first-principles QCD as they occur in the non-perturbative regime, especially at low transverse momentum, $p_T$. In addition, the transport calculations need to be embedded into realistic simulations of the rapidly expanding QCD medium, the dynamics of which vary in the different approaches. 

The fair success of these models in describing HF measurements, given their differing implementation of the HQ dynamics and medium evolution,
translates into significant uncertainties regarding the scientific conclusions that can be obtained from them. In order to gain a better understanding of the relevant commonalities and differences between these models a series of comparative studies have been performed in recent years: For example, in Ref.~\cite{Rapp:2018qla} benchmark HQ interactions within different medium evolutions and different hadronization schemes have been investigated, in Ref.~\cite{Cao:2018ews} different descriptions of the interaction of heavy quarks with the QGP have been confronted, and in Ref.~\cite{Xu:2018gux} the influence of different initial-production mechanisms of heavy quarks and expansion scenarios of the QGP have been analyzed. These works also contain discussions of the theoretical approaches that are being employed in the description of the HQ transport through the QGP medium. Significant commonalities were found between different approaches, thereby improving our understanding of HQ dynamics in a QGP medium. Initial studies of the different hadronization models were also carried out~\cite{Rapp:2018qla}. With this article, we provide a dedicated study on the hadronization process in HF transport models by conducting in-depth comparative studies that have not been done before.   

The paper is organized as follows. In Sec.~\ref{sec_hadro}, we provide a general introduction to hadronization mechanisms for heavy quarks in relativistic heavy-ion collisions, followed by a detailed description of the implementation of these hadronization mechanisms into different dynamical models of HQ transport in Sec.~\ref{sec_models}. In Sec.~\ref{sec_setup}, we define a common environment of an expanding fireball that all models for HQ hadronization are subjected to. Section~\ref{sec_model-comp} provides the results and discussion of systematic comparisons of these models, facilitated by identical HQ distributions at the hadronization hyper-surface as a common input. 
Conclusions are given in Sec.~\ref{sec_concl}. 

%==================================
\section{Hadronization Mechanisms}
\label{sec_hadro}
%===================================
In this section, we provide a brief overview of the hadronization schemes that will be encountered in the model implementations, specifically quark fragmentation typically encoded in universal fragmentation functions in Sec.~\ref{ssec_frag} and quark recombination in Sec.~\ref{ssec_reco}, specifically instantaneous coalescence models (ICMs) based on phase-space Wigner density (Sec.~\ref{sssec_icm}), and the resonance recombination model (RRM) which is carried out in momentum space (Sec.~\ref{sssec_rrm}).

%%%%%%%%%%%%%%%%%%%%%%%%%%%%
\subsection{Fragmentation}
\label{ssec_frag}
%%%%%%%%%%%%%%%%%%%%%%%%%%%%
Independent fragmentation of partons into hadrons is the standard way to describe hadronization in elementary collision systems, such as $pp$ and $e^+e^-$. 
It is based on the assumption that
the differential cross section for a hadron $H$ with momentum $P_H$ factorizes into a hard production cross section for a high-energy parton ($i$) and the fragmentation fragmentation function, ${D}_{i\to H}$,  as~\cite{Owens:1986mp}
\begin{eqnarray}
E\frac{d\sigma_H}{d^3P_H}= E_p \frac{d\sigma_i}{d^3p_i} \otimes \mathcal {D}_{i\to H}(z) \ ,
\label{eq:frag}
\end{eqnarray}  
where $E_i\frac{d\sigma_i}{d^3p_i}$ is the differential perturbative cross section for a parton with momentum $p_i$, and $z=P_H/p_i$ is the momentum fraction carried by the hadron. The symbol $\otimes$ denotes a generic convolution, including the relevant fragmentation fraction $z$ Jacobian. The fragmentation function, $\mathcal{D}_{i\to H}$, is a non-perturbative quantity but it is considered to be universal and usually extracted from experiments such as $e^+e^-$ collisions. There are various choices of fragmentation functions see, e.g., the review in Ref.~\cite{Andronic:2015wma}. The most frequently used fragmentation function for heavy quarks is from Peterson~\cite{Peterson:1982ak}
in which the HF meson carries most of the momentum of the heavy quark $Q$,  
\begin{eqnarray}
\mathcal {D}_{Q\to H}(z) \propto{1\over z[1-{1\over z}-{\epsilon \over 1-z}]^2} \ ,
\end{eqnarray}
where $\epsilon=m_q^2/m_Q^2$, the ratio of the light- and heavy-quark mass squared, is an adjustable parameter. It is commonly fixed by experimental data, e.g., in $e^+e^-$ collisions. 

Equation (\ref{eq:frag}) is considered to be valid if $p_H\gg \Lambda_{QCD}$, i.e., for high-$p_T$ $D$-mesons.
Another popular choice is the HQET-based fragmentation function. For the production of a pseudoscalar meson, $H_P$, or a vector meson, $H_V$ from a heavy quark $Q$, the fragmentation function can be expressed as~\cite{Braaten:1994bz,Cacciari:2005rk} 
\begin{eqnarray}
\mathcal {D}_{Q\to H_P}&\propto&{rz(1-z)^2\over [1-(1-r)z]^6}\Big[ 6-18(1-2r)z 
\nonumber\\
&+&(21-74r+68r^2)z^2
\nonumber\\
&-&2(1-r)(6-19r+18r^2)z^3
\nonumber\\
&+&3(1-r)^2(1-2r+2r^2)z^4\Big],
\nonumber\\
\mathcal {D}_{Q\to H_V}&\propto&{rz(1-z)^2\over [1-(1-r)z]^6}\Big[ 2-2(3-2r)z 
\nonumber\\
&+&3(3-2r+4r^2)z^2
\nonumber\\
&-&2(1-r)(4-r+2r^2)z^3
\nonumber\\
&+&3(1-r)^2(3-2r+2r^2)z^4\Big] \ .
\end{eqnarray}
Here, the adjustable parameter is $r=(m_H-m_Q)/m_H$, which can be interpreted as the ratio of the constituent mass of the light quark to the meson mass. $r$ can also be fixed by experimental data. 

Finally, some groups directly use the fragmentation function provided in PYTHIA with a modified Lund string fragmentation function~\cite{Sjostrand:2006za}. It can be expressed as
 \begin{eqnarray}
\mathcal {D}_{Q\to H}&\propto&{1\over z^{1+rbm_Q^2}}z^{a_\alpha}\left(1-z\over z \right)^{a_\beta}\exp\left(-{bm_T^2\over z} \right),
\label{eq.fpythia}
\end{eqnarray}
where $r$, $a_\alpha$, $a_\beta$, and $b$ are free parameters, and $m_T=\sqrt{m_Q^2+p_T^2}$ is the transverse energy. In PYTHIA, $r=1.32$, $a_\alpha=a_\beta=0.68$, and $b=0.98$ are the default values. This fragmentation function depends separately on the HQ transverse momentum.

%%%%%%%%%%%%%%%%%%%%%%%%%%%%%%%%%
\subsection{Parton Recombination}
\label{ssec_reco}
%%%%%%%%%%%%%%%%%%%%%%%%%%%%%%%%%
Experimental results indicate that in the hot and dense medium, created in relativistic heavy-ion collisions, an additional hadronization mechanism is at work. This was motivated by the enhancement of baryon-to-meson ratios, such as $p/\pi$ and $\Lambda/K$ ratio, observed in heavy-ion collisions at both RHIC~\cite{PHENIX:2003tvk,STAR:2006egk,STAR:2019ank} and the LHC~\cite{ALICE:2015dtd,ALICE:2021bib}, and an approximate quark number scaling of the elliptic flow~\cite{STAR:2003wqp,PHENIX:2006dpn,STAR:2017kkh,ALICE:2014wao}. The parton recombination model was proposed and successfully used to describe these phenomena~\cite{Hwa:2002tu,Greco:2003xt,Fries:2003kq,Molnar:2003ff}, primarily in the regime of intermediate $p_T\lesssim5$ GeV (the high-$p_T$ region is dominated by fragmentation). The most popular phenomenological model for the recombination process is the coalescence model, which assumes that all quarks hadronize at an equal-temperature hypersurface and that the coalescence probability is given by a Wigner function~\cite{Fries:2003kq,Fries:2008hs}. 
Given that HF observables exhibit similar features as the aforementioned light-quark observables (e.g., the $\Lambda_c/D^0$ ratio exhibits an enhancement~\cite{ALICE:2021bib} akin to the $p/\pi$ or $\Lambda/K$ ratio), it is natural to presume that heavy quarks also hadronize via parton recombination in the low- and intermediate-$p_T$ domains.

In the following, we briefly recollect the main features of ICMs (Sec.~\ref{sssec_icm}) and RRM (Sec.~\ref{sssec_rrm}).

%%%%%%%%%%%%%%%%%%%%%
\subsubsection{Instantaneous Coalescence (ICM)}
\label{sssec_icm}
%%%%%%%%%%%%%%%%%%%R
In the sudden coalescence models, the momentum spectrum of the produced hadron $H$ can be written as
\begin{eqnarray}
{dN_{H}\over d^3{\bf P}}&=&g_H\int \prod_{i=1}^N{d^3p_i\over (2\pi)^3E_i}p_i\cdot d\sigma_i F({\bf x}_1...{\bf x}_N,{\bf p}_1...{\bf p}_N)\nonumber\\
&\times&W({\bf x}_1...{\bf x}_N,{\bf p}_1...{\bf p}_N)\delta^{(3)}\left({\bf P}-\sum_{i=1}^N{\bf p}_i\right),
\label{dNcoal}
\end{eqnarray}   
where $g_H$ is a degeneracy factor of color and spin; ${\bf P}$ and ${\bf p}_i$ are the momenta of HF hadron and constituents quarks, respectively. The $\delta$-function ensures the conservation of 3-momentum, $d\sigma_i$ is a hypersurface element, and $F({\bf x}_1...{\bf x}_N,{\bf p}_1...{\bf p}_N)$ represents the constituent-anti-/quark phase-space distribution functions. It amounts to a product of single-particle distribution functions when neglecting correlations between the constituents. The Wigner density of the hadron, $W({\bf x}...{\bf x}_N,{\bf p}_1...{\bf p}_N)$, is used as the recombination probability and can be constructed from the hadron wave function. For the two-body case, it can be defined as
\begin{eqnarray}
W({\bf r}, {\bf p}_r)=\int d^3{\bf y}e^{-i{\bf p}_r\cdot {\bf y}}\psi({\bf r}+{{\bf y}\over 2})\psi^*({\bf r}-{{\bf y}\over 2})  \ ,
\end{eqnarray}  
where ${\bf r}$ and ${\bf p}_r$ are the relative distance and momentum in the center-of-mass (CM) frame of the hadron, respectively.
The wave function $\psi$ can, e.g., be obtained by solving the two-body Schr\"odinger equation. 
In practice, the Wigner density is not used directly. Rather it is assumed that for a hadron in an $S$-wave state, it can be approximated by a Gaussian Wigner density (which can be obtained analytically when using, e.g., the spherical harmonic oscillator) which gives the same root-mean-square (rms) radius as the original Wigner density, 
\begin{eqnarray}
W({\bf r}, {\bf p}_r)=8\exp\left(-{r^2\over \sigma^2}-p_r^2\sigma^2\right) \ .
\end{eqnarray}  
The parameter $\sigma$ in the Wigner function is related to the rms radius of the hadron, $\sigma^2=2\langle r^2 \rangle/3$. Some authors relate the width to the rms charge radius~\cite{Hwang:2001th,Albertus:2003sx} instead of mass radius (see, e.g., in Eq.~\eqref{eq.c1} below). 
%---------------------------------------------------------------------
\begin{figure}[!htb]
\includegraphics[width=0.4\textwidth]{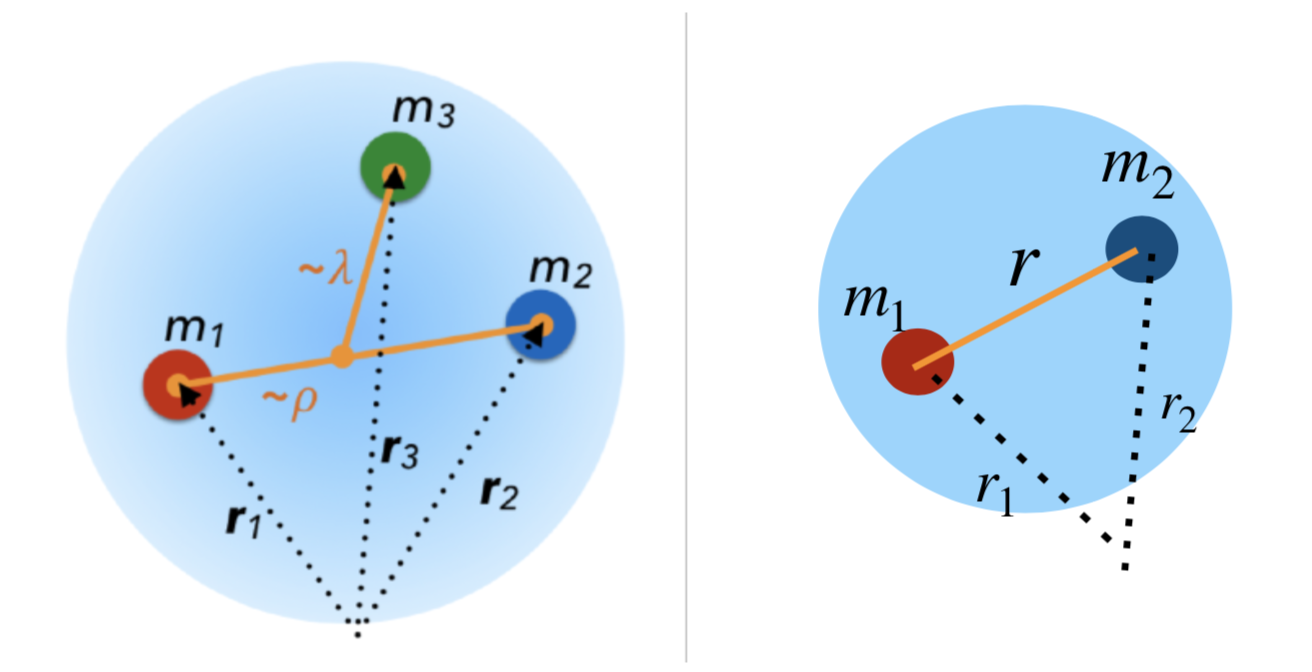}
\caption{Illustration of the relative constituent-quark coordinates for mesons and baryons.}
\label{fig0}
\end{figure}
%---------------------------------------------------------------------

For baryons, the three-body problem can be reduced into two two-body problems by separating the CM motion with the help of Jacobian transformations, for details see Ref.~\cite{Zhao:2020jqu}. There are two relative coordinates, ${\bm \rho}$ and ${\bm \lambda}$ as shown in Fig.~\ref{fig0}. The relative wave function of the three-body system in the ground state is assumed to be factorized as $\Psi=\psi({\bm \rho})\psi({\bm \lambda})$. 
Consequently, the Wigner density can be expressed as a product of two Gaussians,
\begin{eqnarray}
W({{\bm \rho}}, {{\bm\lambda}}, {\bf {p}}_\rho, {\bf {p}}_\lambda)&=&8^2\exp\left(-{\rho^2\over \sigma_\rho^2}-p_\rho^2\sigma_\rho^2\right)\nonumber\\
&&\times\exp\left(-{\lambda^2\over \sigma_\lambda^2}-p_\lambda^2\sigma_\lambda^2\right) \ ,
\end{eqnarray}
where ${\bf p}_\rho$ and ${\bf p}_\lambda$ are the relative momentum corresponding to the relative coordinates, ${\bm \rho}$ and ${\bm \lambda}$. The widths $\sigma_\rho$ and $\sigma_\lambda$ are related to the inner radii via
\begin{eqnarray}
\sigma_\rho^2={2\over3}\langle \rho^2\rangle,\nonumber\\
\sigma_\lambda^2={2\over3}\langle \lambda^2\rangle \ .
\end{eqnarray} 
The rms radius can be obtained by solving the three-body Schr\"odinger equation.

We note that for comparisons to experimental data, one needs to consider excited states as they contribute through decay feeddown to the ground states. For radial and angular excited states, the Wigner function will become complicated but is still manageable by utilizing the wave function of the 3-D isotropic harmonic oscillator. 

The probability density, $dN_H(\mathbf{P},\mathbf{p}_Q)$, to produce a heavy hadron, $H$, with momentum $\mathbf{P}$ from a heavy quark, $Q$, with momentum $\mathbf{p}_Q$ is defined by introducing a delta distribution centered at $\mathbf{p}_Q$ into $F$ in Eq.~\eqref{dNcoal}. One then defines the total recombination probability for this heavy quark to hadronize through coalescence as
\begin{equation}
P_{\rm rec}(\mathbf{p}_Q)= \sum_H \int 
d\mathbf{P}\,dN_H(\mathbf{P},\mathbf{p}_Q).
\label{Preccoal}
\end{equation}
While this quantity is boost invariant, it is not the case for the spectra defined at Eq.~\eqref{dNcoal}, due to the violation of the energy-momentum conservation inherent to the coalescence process. A particular frame thus needs to be chosen, usually taken as the local fluid rest frame.

%%%%%%%%%%%%%%%%%%%%%%%%%%%%%
\subsubsection{Resonance Recombination (RRM)}
\label{sssec_rrm}
%%%%%%%%%%%%%%%%%%%%%%%%%%%%%
The resonance recombination model (RRM) has been derived from a Boltzmann equation where a recombination rate is obtained by utilizing a resonant quark anti-quark cross section (or pertinent scattering matrix element) for hadron fomration~\cite{Ravagli:2007xx}, instead of the Wigner density. In the long time limit, one can formulate this as a coalescence probability while still using off-equilibrium anti-/quark distributions akin to a linear-response theory.
In a similar spirit, a different recombination criterion has recently been proposed~\cite{Beraudo:2022dpz} for HF hadron production in heavy-ion collisions. The main difference between these models is whether the recombination is carried out in momentum space or utilizes phase-space criteria. In particular, the RRM obeys 4-momentum conservation and satisfies the equilibrium limit, i.e., it produces equilibrium hadron distributions if equilibrated quark distributions are used as input (including collective-flow effects). It has been implemented on a hydrodynamic hypersurface in Ref.~\cite{He:2011qa} and extended to HF baryons in Ref.~\cite{He:2019vgs}. 

Let us briefly recall the main steps in deriving the RRM.
For mesons, one starts from a distribution function, $f_M$, that satisfies the Boltzmann equation, 
\begin{eqnarray}
\left({\partial\over \partial t} +{\bf v}\cdot \nabla \right)f_M({\bf R},{\bf P})=-{\Gamma\over \gamma} f_M({\bf R},{\bf P})+\beta({\bf R},{\bf P}) \ ,
\end{eqnarray}  
where ${\bf R}$ and ${\bf P}$ are the position and momentum of the meson,  $\gamma=E_M/m$ is a Lorentz factor with meson mass $m$ and energy $E_M$, and $\Gamma$ is the width of the meson. The first and second terms on the right-hand side are loss and gain terms. In equilibrium, the collisional term disappears, yielding
\begin{eqnarray}
-\int{d^3{\bf R}d^3{\bf P} \over(2\pi)^3}{\Gamma \over \gamma}f_{M}^{eq}+\int{d^3{\bf R}d^3{\bf P} \over(2\pi)^3}\beta =0 \ .
\end{eqnarray}  
The integrated meson yield can then be expressed as
\begin{eqnarray}
N_{M}&=&\int{d^3{\bf R}d^3{\bf P} \over(2\pi)^3}f_{M}^{eq}\nonumber\\
&=&{\gamma\over \Gamma}\int{d^3{\bf R}d^3{\bf P} \over(2\pi)^3}\int{d^3{\bf p}_r \over8(2\pi)^3}f_q({\bf R},{\bf P},{\bf p}_r)f_{\bar q}({\bf R},{\bf P},{\bf p}_r)
\nonumber\\
&&\times \sigma(s)v_{rel} \ ,
\end{eqnarray}
where $f_q$ and $f_{\bar q}$ are the phase space distributions of quark and antiquark, and ${\bf p}_r={\bf p}_1-{\bf p}_2$ and ${\bf P}={\bf p}_1+{\bf p}_2$ are their relative and total momentum,  
and $v_{rel}$ is the relative velocity between quark and antiquark. In its current implementation, the production cross section is assumed to be of a Breit-Wigner form,
\begin{eqnarray}
\sigma(s)=g_\sigma {4\pi\over k^2}{(\Gamma m)^2\over (s-m^2)^2+(\Gamma m)^2} \ ,
\end{eqnarray}  
where $g_\sigma=g_M/(g_qg_{\bar q})$ is the spin-color degeneracy, and $k$ denotes the 3-momentum of the quarks (heavy or light) in the CM frame. $s$ is the invariant mass. For baryons, a two-step process has been employed~\cite{He:2019vgs}, where the first two quarks form a bound diquark, followed by its recombination with another quark into a baryon. 

%==============
\section{Model descriptions}
\label{sec_models}
%===============
While the basic features of hadronization via fragmentation or recombination are well-defined and understood, different models vary in their implementation of the respective hadronization schemes.
In this section, we will provide detailed descriptions of the hadronization schemes used in each model setup. For HQ initial production and energy loss in these models, we refer to the original literature~\cite{vanHees:2005wb,Gossiaux:2009mk,He:2011qa,Cao:2015hia,Rapp:2018qla,Plumari:2017ntm,Cao:2019iqs,Song:2015sfa,Song:2015ykw,He:2019vgs,Li:2020zbk,Beraudo:2022dpz}. Here we limit ourselves to the comparison of those quantities that directly influence the hadronization scheme. These may include:
\begin{enumerate}
\item The functional form of the recombination probability;
\item Parameters such as the relative distance and relative momentum in the recombination probability; 
\item The distribution of light quarks which coalesce with the heavy quark; 
\item The quark masses and widths in the Wigner function;
\item The number of resonance states accounted for;
\item The choice of fragmentation function. 
\end{enumerate}

We need to define and distinguish different Lorentz frames used in the following discussion:
\begin{enumerate}
\item [a)] The CM frame of two colliding nuclei, which is also called the laboratory (Lab) frame;
\item [b)] The fluid rest frame, is usually based on a hydrodynamic description of the expansion of the hot medium; different regions of the medium have different fluid velocities. The fluid rest frame is the stationary frame inside the flowing fluid cell;
\item [c)] The CM frame of the light and heavy quark or the diquark and the heavy quark (in the case of baryons).
\end{enumerate}

As decay feeddowns from excited charm hadrons play an important role in the abundance (and spectra) of the various ground states (in particular, for $D^0$, $D_s$, and $\Lambda_c$) in the following analyses, we have included a table summarizing the excited states accounted for by each model in Tab.~\ref{tab_excited}.

%================

In the remainder of this section, we discuss the hadronization models by the following groups: Catania (Sec.~\ref{sssec_Cata}), Duke (Sec.~\ref{ssec_Duke}), LBT (Sec.~\ref{ssec_Lbt}), Nantes (Sec.~\ref{ssec_Nant}), PHSD (Sec.~\ref{ssec_Phsd}), TAMU (Sec.~\ref{ssec_Tamu}), Torino (Sec.~\ref{Tori}), LANL (Sec.~\ref{ssec_Lanl}), as well as a summary table of the number of excited states involved in each model, and then a discussion of the various fragmentation schemes employed (Sec.~\ref{ssec_frag-models}).

%%%%%%%%%%%%%%%%%%%%%%%
\subsection{CATANIA}
\label{sssec_Cata}
%%%%%%%%%%%%%%%%%%%%%%%
The Catania model adopts a coalescence plus fragmentation scheme for HF hadronization. For the coalescence part, a Gaussian shape in space and momentum for the Wigner function is used (including only $S$-wave states),
\begin{eqnarray}
W(r_i,p_{ri})=\prod_{i=1}^{N_q-1} A_W \exp\left(-{r_{i}^2\over \sigma_{i}^2}-p_{ri}^2\sigma_{i}^2\right) \ ,
\end{eqnarray}  
where $N_q$ is the number of constituent quarks and $A_W$ is a normalization constant that has been fixed to guarantee that in the limit of vanishing charm-quark momentum in the fluid rest frame, $p_c\to0$, all charm quarks hadronize by coalescence into a heavy hadron. This is imposed by requiring that the total coalescence probability gives $\lim_{p\to0}P_{tot}=1$. Going into the CM frame of the particles involved in the process the relative coordinates are defined as follows.
For mesons, they are given by
\begin{eqnarray}
r_{1}=|{\bf x}_1-{\bf x}_2|, ~ p_{r1}={|m_2{\bf p}_1-m_1{\bf p}_2|\over m_1+m_2} \ ,
\label{eq:relmom-ct}
\end{eqnarray}  
while for baryons they are defined as
\begin{eqnarray}
r_{1}={|{\bf x}_1-{\bf x}_2|\over \sqrt{2}}, ~ p_{r1}=\sqrt{2}{|m_2{\bf p}_1-m_1{\bf p}_2|\over m_1+m_2} \ 
\end{eqnarray} 
for the diquark, and $r_{2}, p_{r2}$ for the quark-diquark are given by
\begin{eqnarray}
r_{2}&=&\sqrt{2\over 3}\left|{m_1 {\bf x}_1+m_2 {\bf x}_2\over m_1+m_2}-{\bf x}_3\right|, \nonumber\\
p_{r2}&=&\sqrt{3\over 2}{|m_3({\bf p}_1+{\bf p}_2)-(m_1+m_2){\bf p}_3|\over m_1+m_2+m_3} \ ,
\end{eqnarray}  
as shown in Fig.~\ref{fig0}.

The width in the Wigner function is linked to the size of the hadron and, in particular, to the rms charge radius of the hadron, $\langle r^2\rangle_{ch}=\sum_{i=1}^N\langle (x_i-R_{cm})^2\rangle$ with $N=2,3$ for mesons and baryons, respectively; $R_{cm}$ is the CM coordinate. For mesons one has
\begin{eqnarray}
\langle r^2\rangle_{ch}={3\over 2}{Q_1m_2^2+Q_2m_1^2\over (m_1+m_2)^2}\sigma_{1}^2 \ , 
\label{eq.c1}
\end{eqnarray}
and for baryons
\begin{eqnarray}
\langle r^2\rangle_{ch}&=&{3\over 2}{Q_1m_2^2+Q_2m_1^2\over (m_1+m_2)^2}\sigma_{r1}^2\nonumber\\
&+&{3\over 2}{(Q_1+Q_2)m_3^2+Q_3(m_1+m_2)^2\over (m_1+m_2+m_3)^2}\sigma_{2}^2 \ ,
\label{eq.c2}
\end{eqnarray}
where $Q_i$ is the charge of the $i$-th quark. The charge radius of the hadrons is taken according to the constituent-quark model. In the Catania model one has $\langle r^2\rangle_{ch}=0.184\rm fm^2$ for $D^+$, meaning $\sigma=\sigma_{1}=3.546\rm GeV^{-1}$ for $D$ mesons, $\langle r^2\rangle_{ch}=0.083\rm fm^2$ for $D_s^+$, meaning $\sigma_{1}=2.475\rm GeV^{-1}$ GeV for $D_s$ mesons. 
For $\Lambda_c$, the rms charge radius is taken as $\langle r^2\rangle_{ch}=0.15\rm fm^2$. Actually, for baryons, there is only one free parameter, because
the two widths are related by the oscillator frequency
$\omega$ through the reduced masses, $\mu_i$, via $\sigma_{i}=1/\sqrt{\mu_i \omega}$: 
\begin{eqnarray}
\mu_1&=&{m_1m_2\over m_1+m_2},\nonumber\\
\mu_2&=&{(m_1+m_2)m_3\over m_1+m_2+m_3} \ ,
\end{eqnarray}
where $\mu_1$ is the reduced mass of two-body system and $\mu_2$ is the reduced mass of the two-body system with the third particle. 
The corresponding widths are $\sigma_{1}=5.556\rm GeV^{-1}$ and $\sigma_{2}=2.924\rm GeV^{-1}$.

In the Catania model, all first excited states for $D$ mesons and $\Lambda_c$ are included. For the resonances, the coalescence probability is multiplied by a suppression factor that takes into account the Boltzmann probability to populate an excited state of energy $E+\Delta E$, at  temperature $T$ and a statistical factor, $(m_R/m_G)^{3/2}\times \exp(-(m_R-m_G)/T)$, where $m_R$ is the resonance mass and $m_G$ is the mass of the ground state.

If a heavy quark hadronizes via the coalescence process, one or two light quarks will be sampled on the hadronization hypersurface. The light quarks are assumed to be thermally distributed. Their momentum distribution in the lab frame satisfies
\begin{eqnarray}
f_q={g\tau m_T\over (2\pi)^3}\exp\left(-{\gamma_T(m_T-p_T\cdot \beta_T)\over T}\right) \ ,
\end{eqnarray}  
where $m_T=\sqrt{p_T^2+m_q^2}$, and $\beta_T$ is determined by the velocity field given by the bulk. The factor $g=6$ is the spin-color degeneracy. The presence of gluons in the QGP is taken into account by converting them to quarks and anti-quark pairs according to the flavor compositions;
$\tau$ is the proper time of the hadronization hypersurface. A uniform distribution in coordinate space is assumed. In this model comparison, the quark masses are taken as $m_{u,d}=0.3$ GeV, $m_s=0.38$ GeV, and $m_c=1.5$ GeV.

Heavy quarks that do not hadronize via coalescence are converted to hadrons by fragmentation. Catania uses the Peterson fragmentation with $\epsilon=0.1$ for charmed mesons and $\epsilon=0.02$ for charmed baryons.

For more details, we refer the reader to Refs.~\cite{Plumari:2017ntm,Minissale:2020bif}.

%=============
\subsection{Duke}
\label{ssec_Duke}
%==============
The hadronization model used in the Duke model is also a combination of coalescence and fragmentation. The coalescence probability is given by the momentum-space Wigner function, and for the results shown here only $S$-wave mesons are taken into account. The Wigner function is taken as a Gaussian,
\begin{eqnarray}
W(p_r)=g_h{(2\sqrt{\pi}\sigma)^3\over V}e^{-\sigma^2{p_r}^2} \ ,
\end{eqnarray}  
where $p_r$ is the relative momentum defined in the CM frame,
\begin{eqnarray}
p_r={|E_2{\bf p}_1-E_1{\bf p}_2|\over E_1+E_2} \ .
\label{eq:relmom-duke}
\end{eqnarray}
The width in the Gaussian distribution can be obtained by $\sigma=1/\sqrt{\mu \omega}$ with $\mu$ being the reduced mass of the light and heavy quark. Duke takes $\omega=0.106$ GeV for all charmed hadrons. So, the related widths are, $\sigma(D)=6.23 \rm~GeV^{-1}$ and $\sigma(D_s)=5.22 \rm~GeV^{-1}$. Both ground states and first excited states of $D$ mesons are considered. However, the hadronization implementation utilized for this work~\cite{Cao:2013ita} does not include baryonic states.  Later implementations of the Duke hadronization model~\cite{Cao:2015hia} do include baryons such as the $\Lambda_c$, $\Sigma_c$, and $\Omega_c$. 

The light-quark distribution in the fluid rest frame satisfies
\begin{eqnarray}
f_q={V\over (2\pi)^3}{g_i\over e^{E_i/T}+1} \ ,
\end{eqnarray}  
where $g_i=6$ is the statistical factor. The quark masses are taken as $m_{u,d}=0.3$ GeV, $m_s=0.475$ GeV, and $m_c=1.27$ GeV.
For fragmentation, Duke uses the string fragmentation in PYTHIA 6.4. The main references for Duke's recombination plus fragmentation approach to HQ hadronization are \cite{Cao:2013ita,Cao:2015hia}.

%===========
\subsection{LBT}
\label{ssec_Lbt}
%===========
Hadronization applied in the LBT model is also described by coalescence plus fragmentation. The coalescence probability is given by the momentum space Wigner function, which can be expressed for  $S$-wave and $P$-wave as
\begin{eqnarray}
W_S(p_r)&=&g_h{(2\sqrt{\pi}\sigma)^3\over V}e^{-\sigma^2{p_r}^2} \ ,
\nonumber\\
W_P(p_r)&=&g_h{(2\sqrt{\pi}\sigma)^3\over V}{2\over 3}\sigma^2{p_r}^2e^{-\sigma^2{p_r}^2}  \ ,
\end{eqnarray} 
where the spatial part of the Wigner function has been integrated over coordinate space. The integral will be canceled by the volume factor associated with the momentum distribution functions of light quarks; $g_h$ is the statistical factor; $p_r$ is the relative momentum between the two constituent quarks in the CM frame,
\begin{eqnarray}
p_r={|E_2{\bf p}_1-E_1{\bf p}_2|\over E_1+E_2} \ ,
\label{eq:relmom-LBT}
\end{eqnarray}  
where $E_1$ and $E_2$ are the energies of the quark and antiquark.
Baryons are treated as two two-body systems (produced by recombining two quarks first and then using their CM to recombine with the third one). 
All $S$ and $P$-wave hadron states, allowed by the spin-orbit coupling in a full 3-dimensional calculation, are included, which covers nearly all charmed hadron species ($\Xi_c$ and $\Omega_c$ are also included) as reported by the particle data group (PDG).

The width of the Wigner function is determined by $\sigma=1/\sqrt{\mu \omega}$ with $\mu$ the reduced mass and $\omega$ the oscillator frequency, $\omega=0.24$ GeV, for all charmed hadrons (both $S$ and $P$-wave states), which is tuned so that the total coalescence probability for a zero-momentum charm quark (which cannot fragment) is equal to one when all $S$- and $P$-wave charmed mesons and baryons are included in the calculation. The quark masses are taken as $m_{u,d}=0.3$ GeV, $m_s=0.4$ GeV, $m_c=1.8$ GeV, and the related widths are $\sigma(D)=4.02\rm~GeV^{-1}$ and $\sigma(D_s)=3.57\rm~GeV^{-1}$. For $\Lambda_c$, there are two widths, $\sigma_{1}=4.93\rm~GeV^{-1}$ and $\sigma_{2}=2.87\rm~GeV^{-1}$, corresponding to the diquark system and quark-diquark system, respectively.

Light quarks in the hot medium are thermalized. Their momentum distribution in the fluid rest frame satisfies
\begin{eqnarray}
f_q={V\over (2\pi)^3}{g_i\over e^{E_i/T}+1} \ ,
\end{eqnarray}  
where $g_i=6$ is the statistical factor. 

The traditional instantaneous coalescence model only respects the 3-momentum conservation but violates the energy conservation. A solution of this problem is proposed in Ref.~\cite{Cao:2019iqs} by allowing quarks to combine into an off-shell state of a charmed hadron first and then decay into its on-shell state by emitting a pion. The pion and the final state charmed hadron are generated back-to-back in the rest frame of the off-shell charmed hadron, and then boosted back into the global frame using the velocity of this off-shell state. Since sufficiently large masses are applied for thermal quarks, the kinematics of such decay is usually allowed. In rare cases when the decay is forbidden by kinematics, a photon is emitted instead of a pion.

Heavy quarks that do not hadronize via coalescence are converted to hadrons by fragmentation. The fragmentation process used in the LBT model is PYTHIA 6.4 and the default Peterson fragmentation is used with $\epsilon=0.05$ for charmed hadrons.

For more details on the hadronization model applied to the LBT calculation, please see Ref.~\cite{Cao:2019iqs}.

%===================
\subsection{Nantes}
\label{ssec_Nant}
%===================
 In the Nantes model, heavy quarks hadronize either by coalescence or by fragmentation.
 The recombination probability is evaluated in the thermal restframe and normalized to one for vanishing HQ momentum in that frame. Thus, in the lab frame, it depends on the fluid cell velocity and the orientation of the hadronization hypersurface. The coalescence mechanism is based on the model of Dover~\cite{Dover:1991zn}. 

The coalescence probability is given by
\begin{eqnarray}
&&W(x_Q,x_q,p_Q,p_q)=
\\
&&\exp\left({(x_q-x_Q)^2-[(x_q-x_Q)\cdot u_Q]^2 \over 2R_c^2}\right)F_\Phi(p_Q,p_q) \ ,
\nonumber
\end{eqnarray}  
where $u_Q$ is the four-velocity of the heavy quark and 
\begin{eqnarray}
F_\Phi(p_Q,p_q)=\exp \left( { (p_Q/m_Q-p_q/m_q)^2\over 2\alpha_d^2}\right)
\end{eqnarray}  
with $\alpha_d=0.51$; $R_c$ is fixed for a given $a_d$ from the normalisation condition
\begin{equation}
\frac{R_c}{\sqrt{2\pi}\hbar} =
\frac{m_Q + m_q}{m_Qmq} \times \left( 4\pi \alpha_d^2 K_2(\frac{1}{\alpha_d^2}) e^{\frac{1}{\alpha_d^2}}\right)^{-1/3} \ .
\end{equation}
In coordinate space, one can obtain $\sigma_r=\sqrt{2}R_c$, but in momentum space one has $\sigma_p=1/(\sqrt{2}\alpha_d\mu)$ by definition of the $F_\Phi$, with $\mu$ as the reduced mass.  The widths, $\sigma$, in coordinate and momentum space are therefore not related by the uncertainty relation. The Nantes model is restricted to the hadronization of $c$ (resp. $b$) into $D$ (resp. $B$) mesons and only the ground states are accounted for.

Light quarks are assumed to be thermalized. Their distribution is given by the Boltzmann-J\"uttner distribution in the fluid rest frame,
\begin{eqnarray}
f_q=g_i \exp(-\sqrt{m^2+p^2}/T)
\end{eqnarray}  
with $g_i=6$ the degeneracy factor. The essential change in the new version of the Nantes model is the light-quark mass.
In the old version, $m_{u,d}=0.1$ GeV, $m_c=1.5$ GeV, while in the new version adopted since 2018, $m_{u,d}=0.3$ GeV, $m_c=1.5$ GeV. 

Heavy quarks that do not coalesce form mesons by fragmentation. The fragmentation function used in the Nantes model is the HQET-based fragmentation function~\cite{Braaten:1994bz,Cacciari:2005rk}.

For more details, we refer the reader to 
Ref.~\cite{Gossiaux:2009mk}.

%==============
\subsection{PHSD}
\label{ssec_Phsd}
%==============
PHSD is a kinetic approach and therefore not based on a  hydrodynamical expansion of the QGP. To make it comparable to the other models it has to be modified. For the comparison, the calculation is carried out in a box of size $10\times 10\times10~ \rm fm^3$. The heavy quark is placed at the center of the box and antiquarks are uniformly distributed in the box with a random momentum given by the thermal distribution at $T=0.18$ GeV and with a random mass from the spectral function of antiquark within the dynamical quasi-particle model (DQPM)~\cite{Berrehrah:2013mua,Moreau:2019vhw}, that is, in the fluid rest frame, 
\begin{eqnarray}
\frac{df_{\bar{q}}}{dm}=\frac{2}{\pi}\frac{2m^2\gamma}{(m^2-M^2)^2+(2m\gamma)^2}{g\over e^{E_i/T}+1}
\label{spectral}
\end{eqnarray}  
with $g=6$ the color-spin degeneracy factor; $M$ and $\gamma$ are the pole mass and spectral width of antiquark, respectively.

The coalescence probabilities for the $S$-wave and $P$-wave states are given by
\begin{eqnarray}
W_S(r,p_r)&=&{8(2s+1)\over 36}e^{-{r^2\over \sigma^2}-\sigma^2p_r^2} \ ,
\\
W_P(r,p_r)&=&{2s+1\over 36}\left({16\over 3}{r^2\over \sigma^2}+{16\over 3}\sigma^2p_r^2-8\right) e^{-{r^2\over \sigma^2}-\sigma^2p_r^2} \ ,
\nonumber
\end{eqnarray}  
where $s$ is the spin of meson. The relative distance and momentum in the CM frame are, respectively,
\begin{eqnarray}
r&=&|{\bf r}_1-{\bf r}_2|,
\nonumber\\
p_r&=&{|m_2{\bf p}_1-m_1{\bf p}_2|\over m_1+m_2} \ .
\label{eq:relmom-PHSD}
\end{eqnarray} 
The width $\sigma$ is related to the rms radius of the $D$ meson through
\begin{eqnarray}
\sigma_s^2&=&{4\over 3}{(m_1+m_2)^2\over m_1^2+m_2^2}\langle r_M^2\rangle 
\end{eqnarray} 
for the $S$-wave state and
\begin{eqnarray}
\sigma_p^2&=&{3\over 5}\sigma_s^2
\end{eqnarray} 
for the $P$-wave state, and $\sqrt{\langle r_M^2\rangle}=0.9$ fm for both $S$-wave and $P$-wave states. Different from the charm quark with $m_c=1.5$ GeV, the light quarks have no fixed mass but are off-shell and sampled with the given distribution of Eq.~\eqref{spectral}~\cite{Berrehrah:2013mua,Moreau:2019vhw}. 
The relative momentum coordinates, Eq.~\eqref{eq:relmom-PHSD}, are defined as in the Catania model, Eq.~\eqref{eq:relmom-ct}, while for the Duke and LBT, Eq.s~\eqref{eq:relmom-duke} and~\eqref{eq:relmom-LBT} it appears as a different definition in terms of the energy and not the masses of the recombining quarks; however, we notice that in the center of mass of the pair, the two definitions correspond to the same quantity, being $\bf{p}_1= - \bf{p}_2$.

Only $S$- and $P$-wave mesons are taken into account for coalescence (not taking into account the principle quantum number), while charm baryons such as $\Lambda_c$ are produced only through fragmentation.
After production, a $P$-wave state immediately decays into an $S$-wave state by emitting a pion. The mass of the $P$ state is taken as 2.46 GeV for the $u$ and $d$ sectors and as 2.57 GeV for the $s$ sector. Since the probability is low, the coalescence process is repeated many times until the probability becomes large enough at low $p_T$. It turns out that about ten trials make the coalescence probability similar to those in heavy-ion collisions at RHIC and LHC energies~\cite{Song:2015sfa,Song:2015ykw,Song:2021mvc}.

If the heavy quarks do not hadronize via the coalescence process, they will convert into charmed mesons or $\Lambda_c$ via fragmentation. The Peterson fragmentation function with $\epsilon=0.01$ for charmed mesons is used in the PHSD model.

For more details, we refer the reader to Refs.~\cite{Song:2015sfa,Song:2015ykw}.

%=============
\subsection{TAMU}
\label{ssec_Tamu}
%================
The TAMU model applies a combined recombination and fragmentation approach as well. The former is realized via the resonance recombination model (RRM)~\cite{Ravagli:2007xx}, where the recombination probability for 
the two-body case is controlled by resonance amplitudes and is currently expressed as a relativistic Breit-Wigner cross section,
\begin{eqnarray}
\sigma(s)=g_\sigma {4\pi\over k^2}{(\Gamma m)^2\over (s-m^2)^2+(\Gamma m)^2} \ ,
\end{eqnarray}  
where $g_\sigma=g_M/(g_qg_{\bar q})$ is a statistical weight, $k$  denotes the quark 3-momentum in the CM frame, and $m$ and $\Gamma$ are the mass and resonance width of the meson, respectively.
The recombination probability for a charm quark with momentum $p_c^*$ in the fluid rest frame to form a meson can be expressed as
\begin{eqnarray}
P_{rec}(p_c^*)=P_0\int {d^3p_q \over(2\pi)^3}f_q(p_q){\gamma_M\over \Gamma_M}\sigma(s)v_{rel} \ ,
\label{eq.rrmcoal2}
\end{eqnarray}
and for baryons as
\begin{eqnarray}
P_{rec}(p_c^*)&=&P_0\int {d^3p_{q1}d^3p_{q2} \over(2\pi)^6}f_{q1}(p_{q1})f_{q2}(p_{q2}){\gamma_B\over \Gamma_B}{\gamma_{dq}\over \Gamma_{dq}}\nonumber\\
&\times&\sigma(s_{12})v_{rel}^{12}\sigma(s_{diqc})v_{rel}^{diqc},
\label{eq.rrmcoal3}
\end{eqnarray}
where $f_{q}$ is the thermal light-quark distribution, 
\begin{eqnarray}
f_q={g_i {\rm e}^{-E/T}} \ 
\end{eqnarray}  
with a degeneracy factor $g_i=6$; $v_{rel}^{12}$ and $v_{rel}^{dqc}$ are the relative velocities between the two light quarks, which form a diquark, and between the diquark and the charm quark, respectively (recall that $v_{rel}^{12}=\sqrt{(p^{\mu}_1p_{2\mu})^2-m_1^2m_2^2}/(E_1E_2)$). The light- and heavy-quark masses are taken as $m_{u,d}=0.3$ GeV, $m_{s}=0.4$ GeV and $m_c=1.5$ GeV, and the diquark mass as $m_{ud}=0.7$ GeV.
The resonance widths for meson, diquark, and baryon are taken as $\Gamma_M=0.1$ GeV, $\Gamma_{dq}=0.2$ GeV, and $\Gamma_B=0.3$ GeV; $\gamma_M$, $\gamma_{dq}$, and $\gamma_{B}$ denoted the Lorentz-gamma factors of the meson, diquark, and baryon states, respectively.
An overall parameter $P_0$ is introduced to adjust the recombination probability equal to 1 for a charm quark at rest in the thermal frame.

The RRM in the TAMU model includes a large set of excited states of charmed hadrons that include all states listed by the PDG~\cite{He:2019tik}, but go well beyond that in terms of ``missing'' charm-baryon states as predicted by the relativistic-quark model (RQM)~\cite{Ebert:2011kk} and lattice QCD~\cite{Bazavov:2014yba,Padmanath:2014lvr}. The excited $\Lambda_c$ states and $\Sigma_c$ states are assumed to feed the ground state $\Lambda_c$ with a 100\% branching fraction.

Different from the most recent version of RRM~\cite{He:2019vgs}, where space-momentum correlations (SMCs) between the diffusing quarks and the phase space distributions of the thermal quarks in the hydrodynamically expanding medium are incorporated, the SMCs are neglected for the present study. This means that the RRM is carried out in momentum space only. 

At the end of the diffusion through the QGP, when the charm quark with momentum $p_T^c$ in the lab frame has entered a fluid cell that hadronizes, its momentum is boosted into the fluid rest frame, where it has a momentum $p_c^*$. The recombination probability, $P_{rec}(p_c^*)$, is then calculated via  Eqs.~\eqref{eq.rrmcoal2} and~\eqref{eq.rrmcoal3}, i.e., for a $p_T^c$ we obtain the corresponding recombination probability $P_{rec}(p_c^*)$. Finally, the recombination probability $P_{rec}(p_T^c)$ in the lab frame can be obtained by dividing the sum of probabilities $P_{rec}(p_T^c)$ by the number of incoming particles with $p_T^c$.

Heavy quarks that do not hadronize via RRM are fragmented into HF hadrons using HQET-based fragmentation functions.
The charm-quark fragmentation probability is identified as $1-P_{rec}(p_T^c)$, which is partitioned over all $D$, $D_s$, and $\Lambda_c$ baryons according to their respective fractions determined in $pp$ collisions~\cite{He:2019tik}. 

For more details, we refer the reader to Refs.~\cite{Ravagli:2007xx,He:2011qa,He:2019tik,He:2019vgs}.

%=============
\subsection{Torino}
\label{Tori}
%==============
The hadronization model developed by the Torino group is based on a \emph{local} color neutralization mechanism: once a heavy quark reaches the hadronization hypersurface it undergoes recombination with the closest opposite color charge, forming an excited color-singlet object which will eventually decay into a final charm hadron plus a soft particle allowing for exact four-momentum conservation. The mechanism works as follows.
First, one defines a hadronization hypersurface, identified by the temperature $T_H=155$ MeV. Once a charm quark reaches a fluid cell belonging to such a hypersurface element it is recombined with a thermal particle from the same cell. This thermal particle can be either a (anti-)quark or a (anti-)diquark. Light quarks and diquarks are assumed to be thermally distributed in the local rest frame of the fluid cell. The algorithm consists of the following steps:

1. Randomly extract the medium particle involved in the recombination process, with a statistical weight given by its corresponding thermal density at the temperature $T_H$, namely
\begin{eqnarray}
n=g_sg_I{T_HM^2\over2\pi^2}\sum_{n=1}^\infty {(\pm1)^{n+1}\over n}K_2\left(nM\over T_H \right) \ ,
\end{eqnarray}
where the spin degeneracy $g_s$ and isospin degeneracy $g_I$ are factorized; $M$ is the quark (diquark) mass taken from the default values of PYTHIA 6.4 ($m_{u/d}=0.33$ GeV, $m_s=0.5$ GeV, $m_{(ud)_0}=0.579$ GeV, $m_{(sl)_0}=0.804$ GeV, $m_{(ss)_1}=1.0936$ GeV,...) and the $-/+$ sign refers to fermions/bosons, respectively.

2. Once the particle species is selected, determine its three-momentum in the local rest frame of the fluid cell from a thermal distribution.

3. Boost the four-momentum of the medium particle to the laboratory frame and recombine it with the heavy quark, constructing the cluster $\mathcal {C}$; 

4. Calculate the invariant mass $M_{\mathcal C}$ of the cluster. If the latter is smaller than the mass of the lightest charmed hadron in that given flavor channel, re-sample the medium particle. If not, after defining an intermediate cutoff $M_{max}$ set to $M_{max}=3.8$ GeV, there are two possibilities:
(1) If $M_{\mathcal C}<M_{max}$ the cluster decays isotropically in its own rest frame into two hadrons, which are then boosted back to the laboratory frame. The daughter-charmed hadron will carry the baryon number and strangeness of the parent cluster.
(2) If $M_{\mathcal C}>M_{max}$ the cluster hadronizes via string-fragmentation simulated through PYTHIA 6.4. 

Only ground-state hadrons, like $D^0,D^+,D_s,\Lambda_c,\Xi_c$ and $\Omega_c$, are populated in the model. The resonance decay is roughly encoded in the decay of the parent clusters, following the PDG branching ratios for resonances of mass around $M_{\mathcal C}$ whenever possible. 

For more details, we refer the reader to Refs.~\cite{Beraudo:2022dpz,Beraudo:2023nlq}, where the model is also applied to charmed-hadron production in $pp$ collisions.

%----------------
\begin{table*}[!bt]
\renewcommand\arraystretch{2.0}
\setlength{\tabcolsep}{3.5mm}
\begin{tabular}{c|ccc}
	\toprule[1pt]\toprule[1pt] 
	& $D$ & $D_s$ & $\Lambda_c$  \\
    \bottomrule[1pt]
  Catania & $D^0$, $D^+$, $D^{*0}$, $D^{*+}$ & $D_s$, $D_s^{*+}$ & $\Lambda_c$, $\Lambda_c(2595)$, $\Lambda_c(2625)$, $\Sigma_c(2455)$, $\Sigma_c(2520)$\\
  Duke & $D^0$, $D^+$, $D^{*0}$, $D^{*+}$  & - & - \\
  LBT & All $S$ and $P$-wave $D^0$ and $D^+$ & All $S$ and $P$-wave $D_s$ & All $S$ and $P$-wave $\Lambda_c$ and $\Sigma_c$ \\
  Nantes & $D^0$ & - & - \\
  PHSD & Most $S$ and $P$-wave $D^0$ & Most $S$ and $P$-wave $D_s$ & - \\
  TAMU & PDG & PDG 
  & RQM \\
  Torino & $D^0$ & $D_s$ & $\Lambda_c$ \\
  LANL & $D^0$, $D^+$, $D^{*0}$, $D^{*+}$ & - & - \\
	\bottomrule[1pt]\bottomrule[1pt]
\end{tabular}
\caption{Number of excited states involved in each model. For the Torino model, the contribution of excited states is encoded in the decay of the clusters.}
\label{tab_excited}
\end{table*}
%----------------

%=============
\subsection{LANL}
\label{ssec_Lanl}
%=============
The LANL model only includes fragmentation processes.
In $pp$ collisions, heavy quarks fragment into HF hadrons which is described by the HQET-based fragmentation function~\cite{Braaten:1994bz,Cacciari:2005rk}. 

However, in high-energy heavy-ion collisions, such as the ones at the LHC, the gluon contribution to heavy-meson production can be as large as $\sim50\%$~\cite{Anderle:2017cgl}. Contributions to parton energy loss and in-medium parton showers in general, to hadron production, can be accounted for in different but related ways~\cite{Kang:2016ofv}. In the LANL model an effective modification of the fragmentation function~\cite{Vitev:2002pf,Vitev:2005he} is employed as
\begin{eqnarray}
\mathcal{D}_{H/c}(z)&\Rightarrow&\int_0^{1-z}d\epsilon P(\epsilon){1\over 1-\epsilon}D_{H/c}\left({z\over 1-\epsilon}\right)\nonumber\\
&+&\int_{z}^1d\epsilon{dN_g\over d\epsilon}(\epsilon){1\over \epsilon}D_{H/g}({z\over \epsilon}) \ ,
\end{eqnarray}
where $P(\epsilon)$ is the probability that the heavy quark loses a fraction of its energy, $\epsilon$, due to multiple gluon emissions, and $dN_g/d\epsilon$ is the distribution of the average gluons as a function of $\epsilon=\omega/E$; $D_{H/c}$ is the fragmentation function of the charm quark into charmed hadrons, whereas  $D_{H/g}$ is the fragmentation function of gluons into charm hadrons. For more details, please see Refs.~\cite{Anderle:2017cgl,Li:2020zbk}.

%==============================
\subsection{Fragmentation Function Comparison}
\label{ssec_frag-models}
\begin{figure}[!htb]
\includegraphics[width=0.4\textwidth]{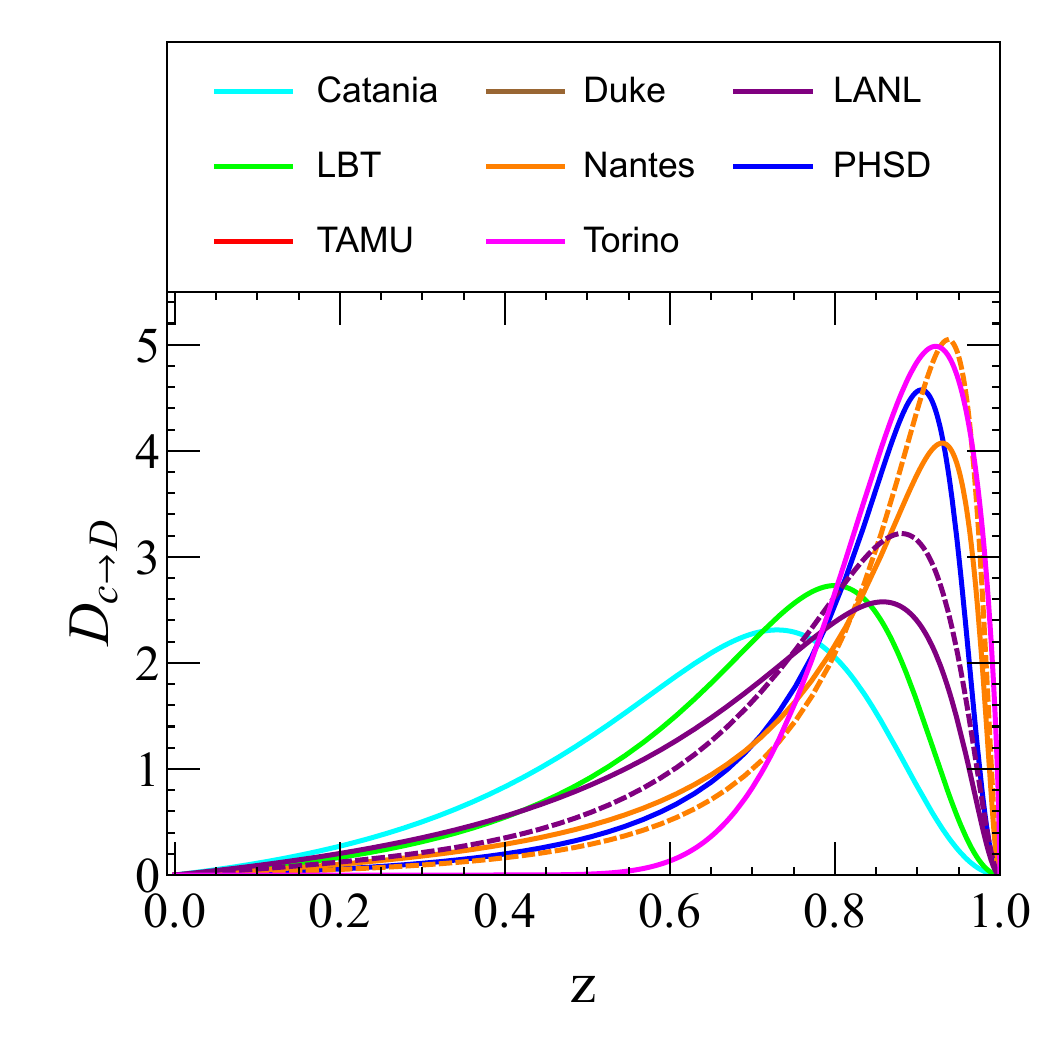}
\caption{The normalized fragmentation function ($c\to D$) used by the different groups. The Duke and Torino group use the fragmentation function of PYTHIA 6.4, which is $p_T$ dependent, as shown in Eq.~\ref{eq.fpythia}; we set $p_T=3$ GeV and $m_c=1.5$ GeV for the plot. Here the solid lines in the LANL and Nantes models are pseudoscalar channels, and dashed lines are vector channels.
}
\label{fig_frag}
\end{figure}

%===============================
Different charm-quark fragmentation functions as used in the approaches described above are compared in Fig.~\ref{fig_frag}. The Catania (light blue), LBT (green), and PHSD (dark blue) groups use the Peterson fragmentation function~\cite{Peterson:1982ak} with the parameter $\epsilon=0.1,0.05,0.01$, respectively.
The Nantes (orange), TAMU (orange), and LANL (purple) groups use HQET-based fragmentation functions.
The TAMU model uses one parameter for the pseudoscalar meson channel, $r=0.1$ for the $D^0$ meson, while resonance states are obtained by the scaling law $r_M/r_{D^0}=((m_M-m_c)/m_{M})/((m_{D^0}-m_c)/m_{D^0})$. In the Nantes model, the charm quark is randomly chosen to fragment via the pseudoscalar or the vector channel with a branching ratio of 0.418. The $r$ parameter is set to $r=0.1$ as well. The LANL model uses fragmentation into scalar and vector charm mesons with the $r$ parameter set to $r=0.2$.
The Duke and Torino groups directly utilize PYTHIA 6.4 for fragmentation.
 
We observe that the fragmentation functions differ considerably in the different approaches. Four of them (Nantes, PHSD, TAMU, and Torino) are rather narrowly peaked around 0.9 and therefore transfer a large momentum fraction to the $D$ mesons, whereas three others (LANL, LBT, and Catania) show a rather broad distribution with a significant probability to produce $D$ mesons which carry half or less of the momentum of the $c$ quark. The ordering is not so much governed by the formalism but rather by the parameters used in each formalism.
%---------------------------------------------------------------------
%---------------------------------------------------------------------

%==================================
\section{Setup of the Model Comparison}
\label{sec_setup}
%=================================
To scrutinize the hadronization mechanisms realized in different evolution frameworks, we embed the different model implementations into an otherwise identical environment.   
In the present study, the expanding QGP is modeled by a simple fireball model~\cite{vanHees:2005wb}. The fireball is cylindrically oriented along the $z$-direction with an elliptic cross sectional area. The long ($a$) and short ($b$) semi-axes of the ellipse are oriented, respectively, along $y$- and $x$-direction and are given by
\begin{eqnarray}
a(t)=a_0&+&v_\infty \left[t-{1-\exp{(-At)}\over A}\right]\nonumber\\
&-&\Delta v\left[t-{1-\exp{(-Bt)}\over B}\right],\nonumber\\
b(t)=b_0&+&v_\infty \left[t-{1-\exp{(-At)}\over A}\right]\nonumber\\
&+&\Delta v\left[t-{1-\exp{(-Bt)}\over B}\right] \ .
\end{eqnarray}
The velocity of the fireball surface along the long and short axis is given by $v_a(t)=\dot a(t)$ and $v_b(t) =\dot b(t)$. The parameters are taken as, $a_0=5.562$ fm, $b_0=4.450$ fm, $v_\infty=0.52$, $\Delta v=0.122$, $A=0.55/\rm fm$, and $B=1.3/ \rm fm$ for Pb-Pb collisions at $\sqrt{s_{\rm NN}}=2.76$ TeV and impact parameter $b=7$ fm, resulting in $v_b(t)>v_a(t)$.

Hadronization is assumed to occur on a hypersurface $R_\mu=(t, {\bf R})$, which is a 4-dimensional surface with equal temperature $T(R_\mu) = T_H$. This corresponds to a hadronization time $\tau=\tau(T_c, {\bf R})$ in the above set-up, which amounts to $t$ =4.86~fm/c at a final temperature of $T_H=180$~MeV. This defines the coordinates and velocities of the fluid cell at the hadronization hypersurface. The chosen temperature value, $T_H$, is somewhat larger than the pseudo-critical temperature of the chiral transition (around $T_{\rm pc}\simeq 155$~MeV)~\cite{Bazavov:2014yba}, but the latter does not have to coincide with the (onset) of hadronization, which is expected to be a continuous process over a finite temperature interval. For the comparative study carried out here, this is of minor importance, where we rather need a fireball with a simple and well-defined flow field (and do not aim at quantitative comparisons to data).

\begin{figure}[!thb]
\includegraphics[width=0.4\textwidth]{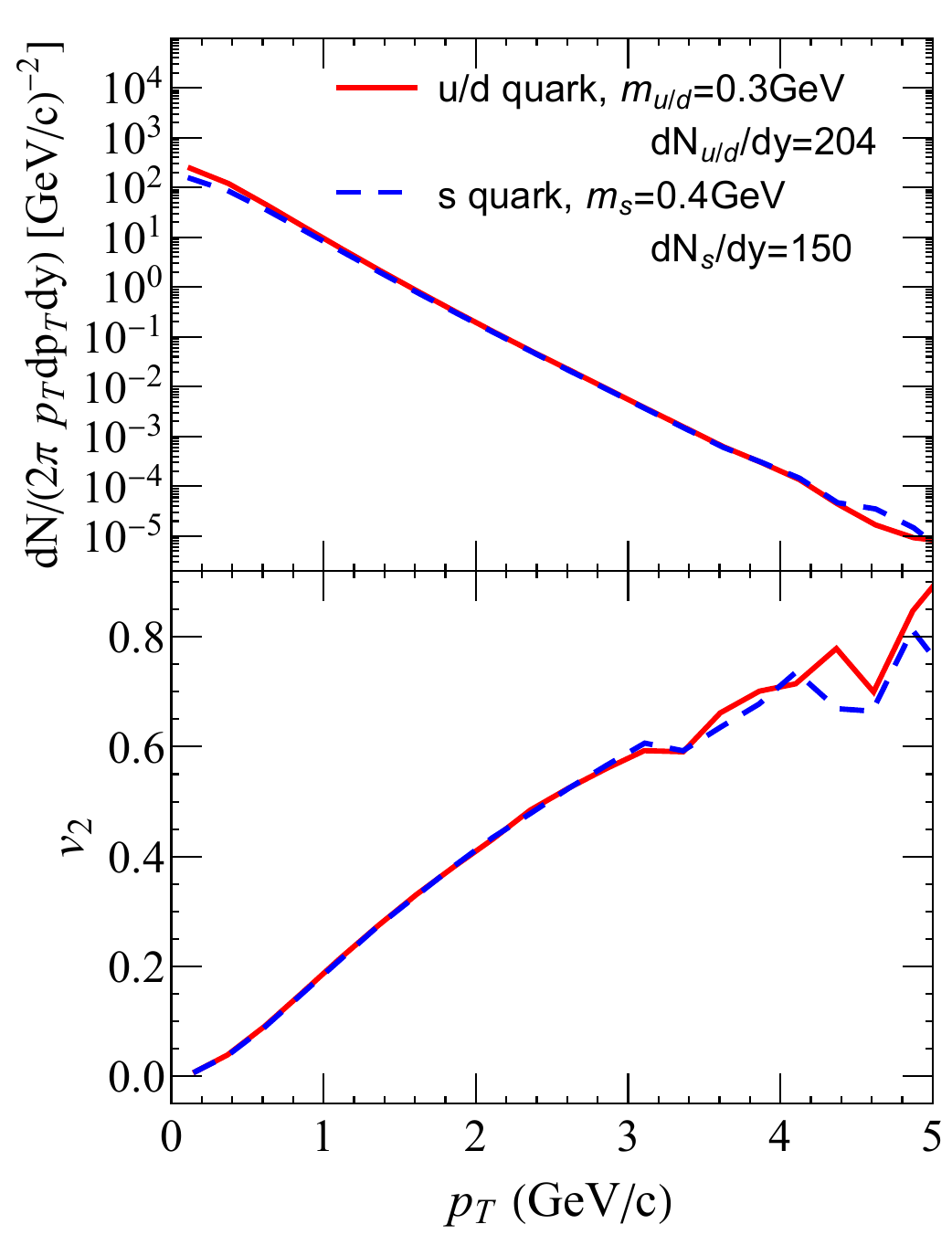}
\caption{The transverse-momentum spectrum  (upper panel) and elliptic flow, $v_2$ (lower panel), of light quarks, $u/d$ and $s$, at the hadronization hypersurface in the lab frame.}
\label{fig.light}
\end{figure}

The light quarks are assumed to be thermalized in almost all models. By taking a thermal distribution for $u$, $d$, and $s$ quarks, their three momenta can be sampled in the fluid rest frame. Due to the different velocities in the $x$- and $y$-directions, the light quarks boosted to the lab frame will carry a non-zero elliptic flow, which is defined as the second-order coefficient of the Fourier expansion of the azimuthal distribution.
The resulting transverse-momentum distribution and the elliptic flow of light quarks are displayed in Fig.~\ref{fig.light}.
%---------------------------------------------------------------------
%---------------------------------------------------------------------------

\begin{figure}[!thb]
\includegraphics[width=0.4\textwidth]{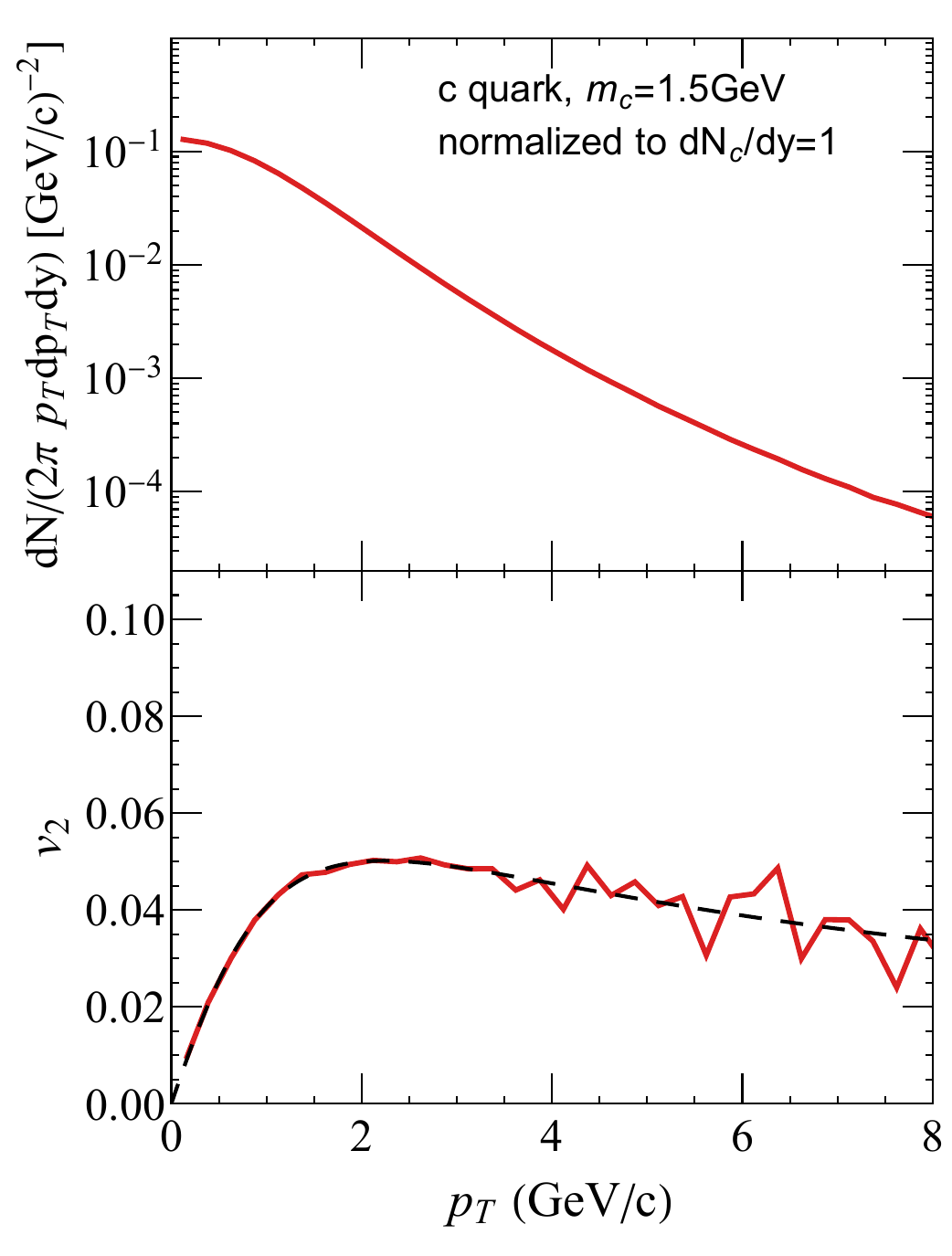}
\caption{The transverse-momentum spectrum in the Lab frame (upper panel) and elliptic flow, $v_2$, (lower panel) of charm quarks on the hadronization hypersurface. The black dashed line shows the $v_2$ from the parameterized formula.}
\label{fig.charm}
\end{figure}

Charm quarks are also assumed to be uniformly distributed on the hadronization hypersurface, and their momentum distribution in the lab frame is taken from the EMMI rapid reaction task framework~\cite{Rapp:2018qla}, displayed in the top panel of Fig.~\ref{fig.charm}. The distribution can be parameterized as $dN/dp_T=53.15(p_T/\mathrm{GeV})/(4.423+(p_T/\mathrm{GeV})^2)^{2.808}$.
The corresponding charm-quark elliptic flow, which is shown in the bottom panel of Fig.~\ref{fig.charm}, will be specified below.

In realistic simulations of heavy-ion reactions, charm quarks move outward through a hypersurface element and typically possess a velocity that is correlated with the one of the local fluid environment, resulting in space-momentum correlations.
In this study, we neglect these correlations: charm quarks are sampled isotropically in each fluid cell at the hypersurface.

%---------------------------------------------------------------------
%---------------------------------------------------------------------

In such prepared environment, the different groups carried out four different calculations:
\begin{enumerate}
\item
The yield of different charm hadrons considering
%----------
\begin{enumerate}
\item only fragmentation (assuming that all heavy hadrons are created by fragmentation); 
\item only coalescence/recombination (all hadrons are produced by coalescence/recombination);
\item the full framework of fragmentation plus coalescence, as realized in the respective model.
\end{enumerate}
%---------
For each case, provide a ``hadronization ratio"~\cite{Rapp:2018qla}
\begin{eqnarray}
H_{AA}(D)&=&{dN_D/dp_T\over dN_c/dp_T},\nonumber\\
H_{AA}(D_s)&=&{dN_{D_s}/dp_T\over dN_c/dp_T},\nonumber\\
H_{AA}(\Lambda_c)&=&{dN_{\Lambda_c}/dp_T\over dN_c/dp_T},
\label{eq:def}
\end{eqnarray}
where $D$ includes $D^0$ and $D^+$. $D_s$ and $\Lambda_c$ are the prompt yields (direct + strong decay). 
%----------------------
\item The elliptic flow $v_2$ of all charmed hadrons with an anisotropic charm-quark spectrum according to
\begin{eqnarray}
{dN_c\over d\vec{p}_T}\Rightarrow{dN_c\over dp_T} \times(1+2v_2(p_T)\cos(2\phi)) \ , 
\label{eq:v2c}
\end{eqnarray}
with $v_2(p_T)=0.0577 p_T/(1+0.4212p_T^{1.6367})$, which is shown as the black dashed line in Fig.~\ref{fig.charm} ($p_T$ in units of GeV).\\

The calculation should consider two cases:
\begin{enumerate}
\item only-fragmentation (assuming all $c$ quarks to hadronize through fragmentation); 
\item the full hadronization scheme used by the respective model (recombination + fragmentation).
\end{enumerate}
%-------------
\item The $H_{AA}$ and $v_2$ of directly produced  $D^0$, $D_s$, and $\Lambda_c$ (without feeddown contributions).

%-----------
\item ${dN(D^0)/ dp_T}$ of direct $D^0$ mesons (without feeddown contributions) produced by a $c$-quark with $p_T^c$ = 3 and 10~GeV.
\end{enumerate}

For the subsequent discussion we define the yield for a given charm hadron with transverse momentum, $P_T$, originating from a $c$-quark with momentum $\mathbf{p}_T^c$ as 
\begin{equation}
dN_H^{\rm mixed}(\mathbf{P}_T,\mathbf{p}_T^c)=
dN_H(\mathbf{P}_T,\mathbf{p}_T^c)+
 (1-P_{\rm rec}(\mathbf{p}_T^c)) \mathcal{D}_{c\rightarrow H},
\label{eq.mixcoal} 
\end{equation}
where $dN_H$ encodes the physics of the recombination. This suggests to organize a model comparison in terms of the 3 ingredients $P_{\rm rec}$, $dN_H$ and $ \mathcal{D}_{c\rightarrow H}$, which is part of the motivation for the tasks to provide not only the ``mixed" (or total) hadronization results but also  ``only-fragmentation'' and ``only-recombination'' cases.

%---------------------------------------------------------------------
\begin{table*}
	\renewcommand\arraystretch{2.0}
	\setlength{\tabcolsep}{3.5mm}
	\begin{tabular}{c||c|c|c||c|c|c||c|c|c}
		\toprule[1pt]\toprule[1pt] 
		\multicolumn{1}{c||}{} & \multicolumn{3}{c||}{$\text{only-fragmentation}$} &   \multicolumn{3}{c||}{$\text{only-recombination}$}&  \multicolumn{3}{c}{$\text{mixed-hadronization}$} \tabularnewline
		\midrule[1.5pt]
		 & \multicolumn{1}{c|}{$D$} & \multicolumn{1}{c|}{$D_s$} &\multicolumn{1}{c||}{$\Lambda_c$} & \multicolumn{1}{c|}{$D$} & \multicolumn{1}{c|}{$D_s$} & \multicolumn{1}{c||}{$\Lambda_c$} & \multicolumn{1}{c|}{$D$} & \multicolumn{1}{c|}{$D_s$} & \multicolumn{1}{c}{$\Lambda_c$}  \tabularnewline
		\midrule[1.5pt]
		Catania & 78.3\% & 8.0\% & 13.7\% & - &  - &  -  & 48.8\% & 6.8\% & 24.3\% \tabularnewline
		\midrule[1pt]
		Duke &  100\% & - & - & 100\% &  -  &  -  & 100\% & - & - \tabularnewline
		\midrule[1pt]
		LBT & 37.8\% & 5.4\% & 3\% & 50.3\% &  14.6\%  &  20.8\%  & 54.7\% & 12.1\% & 15.3\% \tabularnewline
		\midrule[1pt]
		Nantes & 100\% & - & -  & 100\% &  -  &  -  & 100\% & - & - \tabularnewline
		\midrule[1pt]
		PHSD & 81\% & 10\% & - & 67\% &  33\%  &  -  & 75\% & 20\% & - \tabularnewline
		\midrule[1pt]
		TAMU & 60.7\% & 11.5\% & 24.1\% & - &  -  &  -  & 50.2\% & 16.2\% & 22.8\% \tabularnewline
		\midrule[1pt]
		Torino & - & - & - & - &  -  &  -  &50.6\% & 17.9\% & 20.4\% \tabularnewline
		\midrule[1pt]
		LANL & 77.8\% & 10\% & 11.9\% & - &  -  &  -  &- & - & - \tabularnewline
		\bottomrule[1pt]\bottomrule[1pt]
	\end{tabular}
	\caption{Percentage distributions of 3 different ground-state hadron species to which a $c$-quark hadronizes. Presented are the results for only fragmentation (left 3 columns), only recombination (middle 3 columns), and mixed hadronization (right 3 columns).}
 \label{tab_frac}
\end{table*}
%--------------------------------------------------------------------

%---------------------------------------------------------------------
\begin{figure}[!htb]
\includegraphics[width=0.4\textwidth]{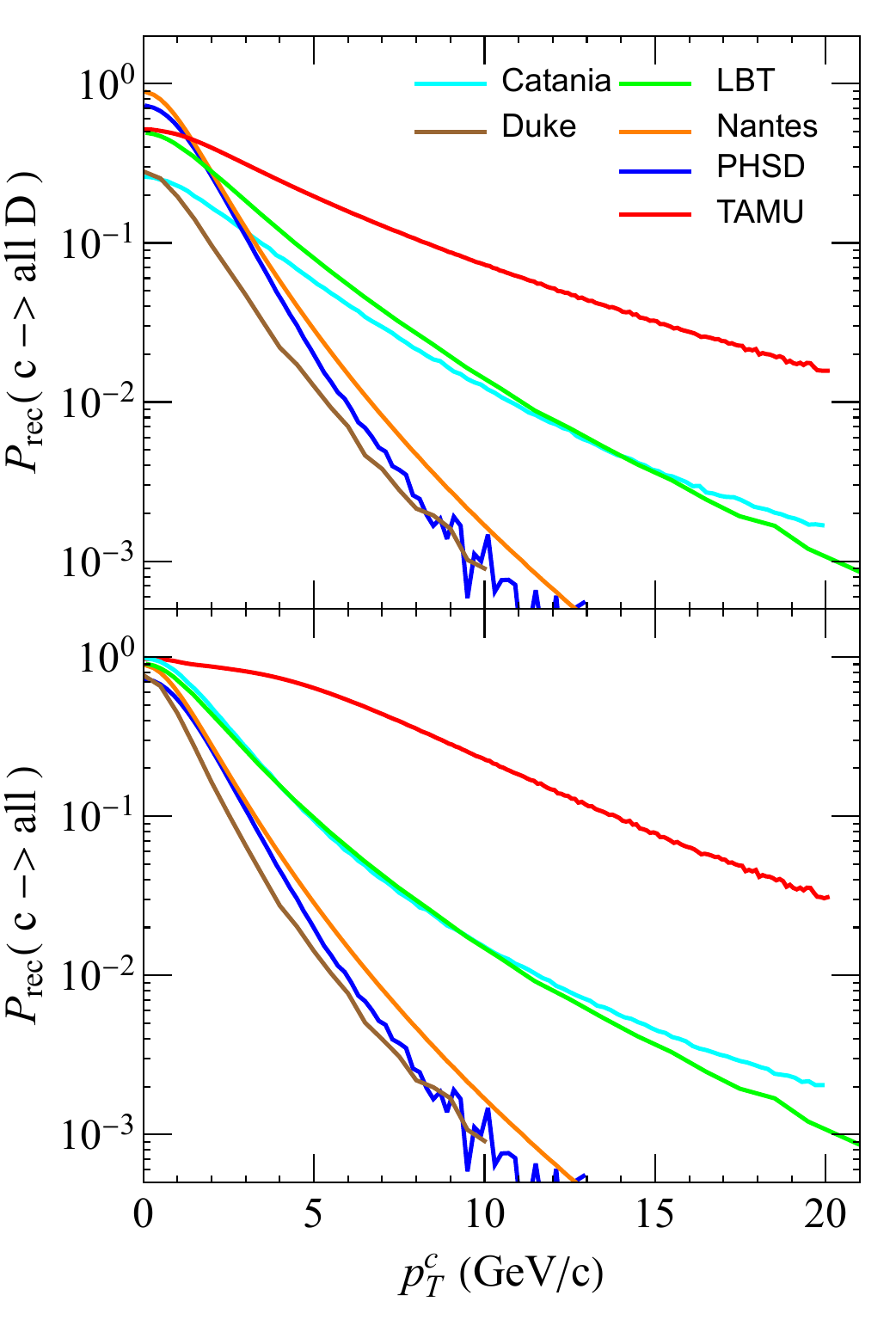}
\caption{The recombination probability of a charm quark of transverse momentum $p_T^c$ in the lab frame, into all charmed mesons (upper panel) and into all charmed hadrons (lower panel).
}
\label{fig1}
\end{figure}
%---------------------------------------------------------------------
%---------------------------------------------------------------------
\begin{figure*}[!htb]
\includegraphics[width=0.8\textwidth]{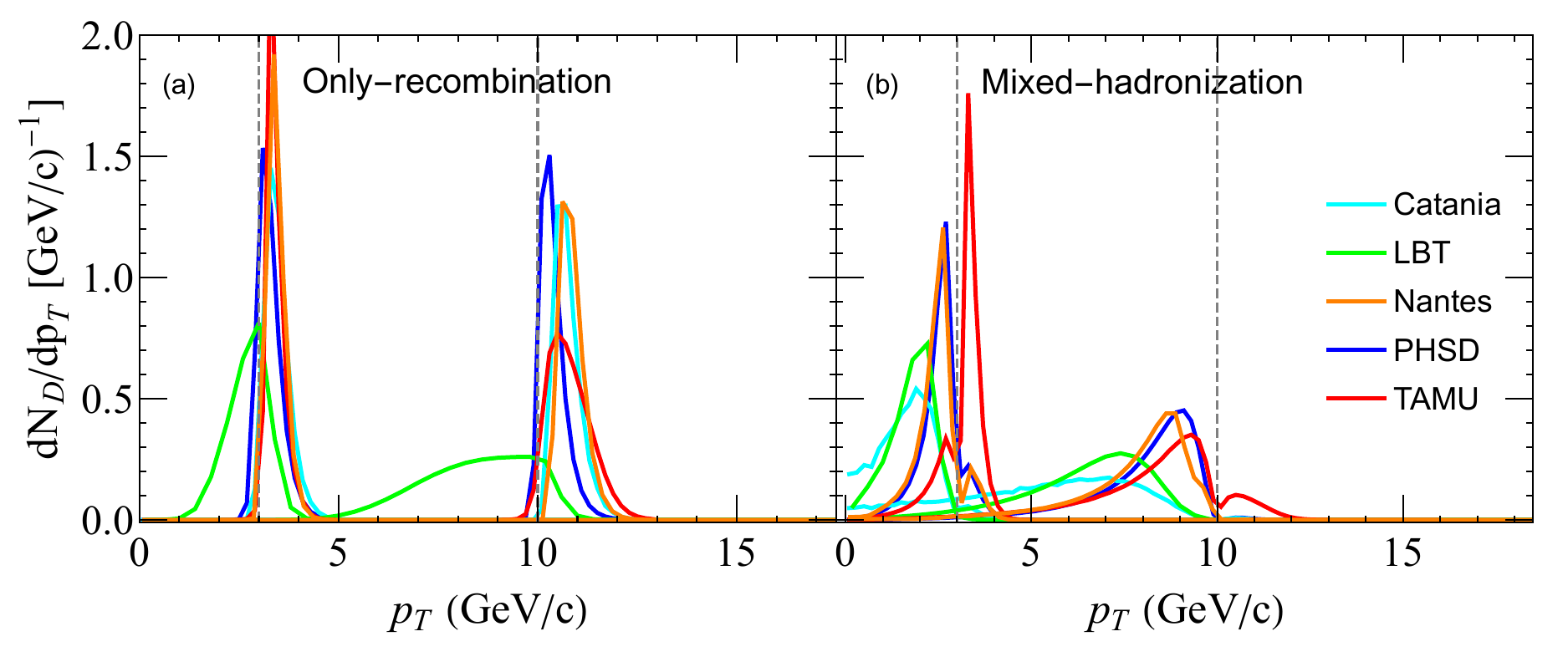}
\caption{The normalized spectrum, $dN_D/dp_T$, of direct $D^0$'s resulting from $c$-quarks with fixed initial transverse momentum $p_T^c$=3.0, 10.0 GeV (indicated by the gray-dashed vertical lines). The only-coalescence and mixed-hadronization processes are plotted in the left and right panel, respectively.}
\label{fig_shift}
\end{figure*}
%---------------------------------------------------------------------

%========================
\section{Results of the Model Comparisons}
\label{sec_model-comp}
%========================
The comparisons laid out in the previous section are carried out in this section and organized as follows. In Sec.~\ref{ssec_prob} we first collect the inclusive production fractions of the ground-state hadrons $D$, $D_s$, and $\Lambda_c$, for the cases where only fragmentation or only recombination is accounted for, and for the combined scenario in each approach, followed by a study of the $p_T$-differential recombination probabilities. In Sec.~\ref{ssec_hadron-pT} we scrutinize the $p_T$ dependence of charm-hadron production using the hadronization ratio, $H_{AA}$~\cite{Rapp:2018qla} and variants thereof, as well as ratios of different hadron species (hadro-chemistry). In Sec.~\ref{ssec_hadron-v2} we analyze how the $p_T$-dependent elliptic flow of the input charm-quark distribution is converted into that of various hadrons.

%%%%%%%%%%%%%%%%%%%%
\subsection{Charm-Hadron Production Fractions}
\label{ssec_prob}
%%%%%%%%%%%%%%%%%%%%
We start our analysis by listing the $p_T$-integrated production fractions of the ground-state hadrons in Tab.~\ref{tab_frac}, i.e., $D$-mesons as well as those carrying an additional conserved quantum number, i.e., the charm-strange $D_s$ meson and the $\Lambda_c$ baryon. We do that for 3 scenarios where all charm quarks are forced to hadronize via either fragmentation or recombination, or in the ``realistic" scenario (i.e., as used in practice) of both mechanisms contributing. Not all entries are filled, especially for the "recombination-only" case, as it is not straightforward to enforce 100\% recombination in several models. Also note that even in the mixed scenario the fractions do not necessarily add up to one, indicating that other charm hadrons with additional conserved quantum numbers contribute (such as charm-strange or doubly-charm baryons).

Next, we turn to the recombination probability, $P_{rec}(p_T^c)$, as a function of charm-quark transverse momentum, $p_T^c$, 
shown in Fig.~\ref{fig1}. This represents the probability, calculated in the local thermal rest frame for most models (but displayed as a function of the charm-lab momentum, $p_T^c$)
that a charm quark hadronizes via recombination.  In Fig.~\ref{fig1}, top,  $P_{rec}$ includes hadronization to all charmed mesons (such as $D^0$, $D^+$, $D_s$, and all excited states), while in the bottom panel we also include charmed baryons (such as $\Lambda_c$, $\Xi_c$, $\Omega_c$, and all excited states).  The color coding is the same as that in Fig.~\ref{fig_frag}. The fragmentation probability $P_{frag}(p_T^c)$ is given by $1-P_{rec}(p_T^c)$.

There is no quantitative guidance from QCD on how $P_{rec}$ should behave as a function of $p_T^c$. Only at large $p_T^c$, the domain of DGLAP jet evolution, hadronization is dominated by universal fragmentation functions and therefore $P_{rec} \to 0$. This lack of guidance is a main reason why the 
$p_T^c$ dependence of $P_{rec}$ is rather different in the different approaches. 
In some approaches fragmentation also contributes at very small $p_T^c$, whereas in others hadronization at small $p_T^c$ is exclusively due to recombination. The latter is based on the assumption that the probability goes to one when $p_T^c\to 0$ (usually applied in the thermal rest frame), whereas in the former case the analytical recombination equations are simply extended to $p_T^c=0$, which implies a significant dependence on the parameters in the respective coalescence model. One should also recall that most coalescence models are performed in the thermal rest frame, while others are directly applied in the lab frame which can cause issues with Lorentz invariance. In fact, also fragmentation functions, usually evaluated in the lab frame, are not guaranteed to be Lorentz invariant.
While the $p_T^c$ dependence for relative contributions from recombination and fragmentation covers a large variation in the models, we nevertheless observe 3 groups. The hardest dependence is found for the TAMU model, likely due to the large set of excited resonances that go out to rather large masses, thus facilitating charm-quark recombination at relatively large momenta and further augmented by the Lorentz boost due to the collective fireball flow.
The next hardest distributions are from LBT and Catania; for the former, off-shell production (with subsequent pion decay) is presumably the dominant mechanism at high $p_T^c$ (akin to the excited states in the TAMU model), while the Catania model features rather small wave function radii compared to the other ICMs (Duke, Nantes, and PHSD), which form the class with the softest $p_T^c$ dependence and show rather good agreement.

%%%%%%%%%%%%%%%%%%%%%%%%%%%%%%%%%
\subsection{Transverse-Momentum Distributions of Charm Hadrons}
\label{ssec_hadron-pT}
%%%%%%%%%%%%%%%%%%%%%%%%%%%%%%%%%%
We first carry out task no.~4 by studying how the transverse-momentum  ($p_T^D$) distribution of $D$-mesons looks like in the different models if one hadronizes a $c$-quark of a given transverse momentum, $p_T^c$.  If created by a fragmentation process, the $D$-meson carries only a fraction of the $p_T^c$ of the $c$-quark, as shown in Fig.~\ref{fig_frag}. For recombination one generally expects the opposite trend, as the momentum of the $D$-meson is the (vector) sum of the HQ and light-quark momenta, and most algorithms favor an alignment of the light-quark $p_T$  with $p_T^c$, especially for ground-state hadrons. The distributions of $p_T^D$ for a charm quark with $p_T^c=3$ and 10\, GeV are displayed in Fig.~\ref{fig_shift} left for the case that all $D$-mesons are created via recombination. To minimize the influence of parameter choices, we have decided to evaluate quantities pertaining to the directly produced hadrons -- hence Figs.~\ref{fig_shift}, \ref{fig7}, and  \ref{fig8} --
requiring the charm- and light-quark masses in the calculations to be $m_c$ = 1.5 GeV, $m_{u/d}$ = 0.3 GeV, $m_s$ = 0.4 GeV. In addition, we also impose $\sigma$ = 0.5 fm (charmed mesons) and  $\sigma_\rho$ = $\sigma_\lambda$ = 0.5\,fm (charmed baryons) for all models that use the Wigner function as recombination probability.

For recombination, almost all models show a slight shift of the maximum towards a larger $D$-meson momentum, $p_T^D$, as compared to $p_T^c$, with a width of the order of 1 GeV for $p_T^c$ = 3 GeV, which increases with $p_T^c$, cf.~right panel of Fig.~\ref{fig_shift}. Only in the LBT model, the produced charm hadron, which is usually off-shell at first and subsequently decays into a pion and an on-shell charm hadron, one has $p_T^D<p_T^c+p_T^q$. If we weight fragmentation and recombination according to Fig.~\ref{fig1}, the form and the maximum of the distributions become rather different, even for $p_T^c$ as low as 3 GeV, see the right panel of Fig.~\ref{fig_shift}. For $p^c_T=10$ GeV most mesons are produced by fragmentation and therefore all models besides the TAMU model show a shift exclusively to lower $D$-meson $p_T$ (while the recombination portion in the TAMU model still generates a component above the $c$-quark momentum). For $p^c_T=3$ GeV we see that contributions from both recombination and fragmentation are operative. 

%---------------------------------------------------------------------
\begin{figure*}[!htb]
\includegraphics[width=0.9\textwidth]{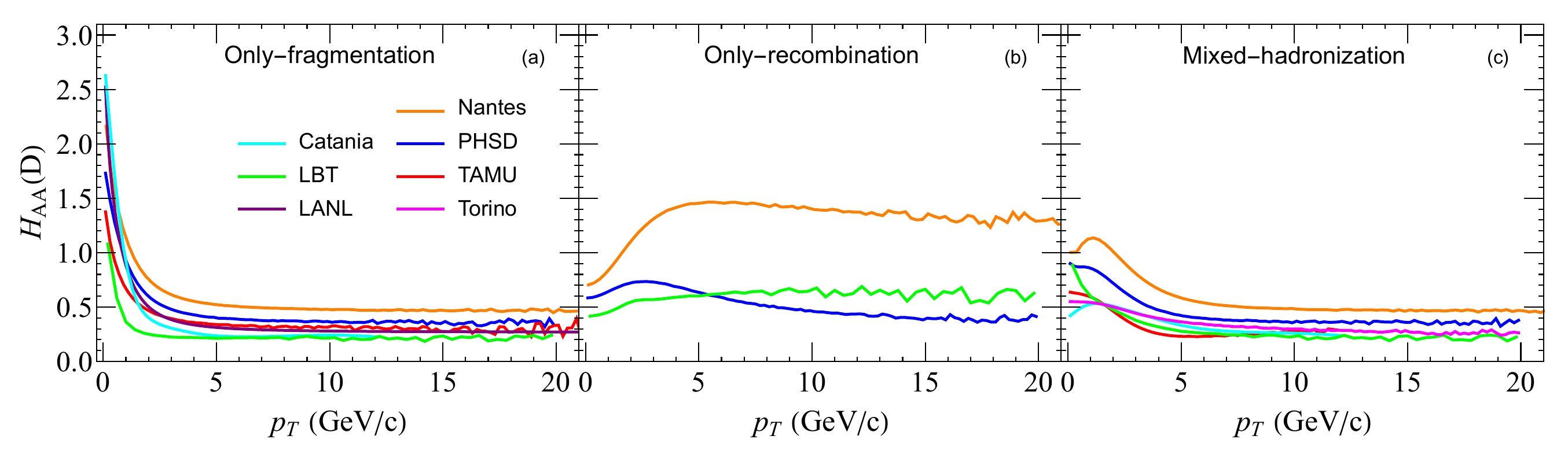}\\
\includegraphics[width=0.9\textwidth]{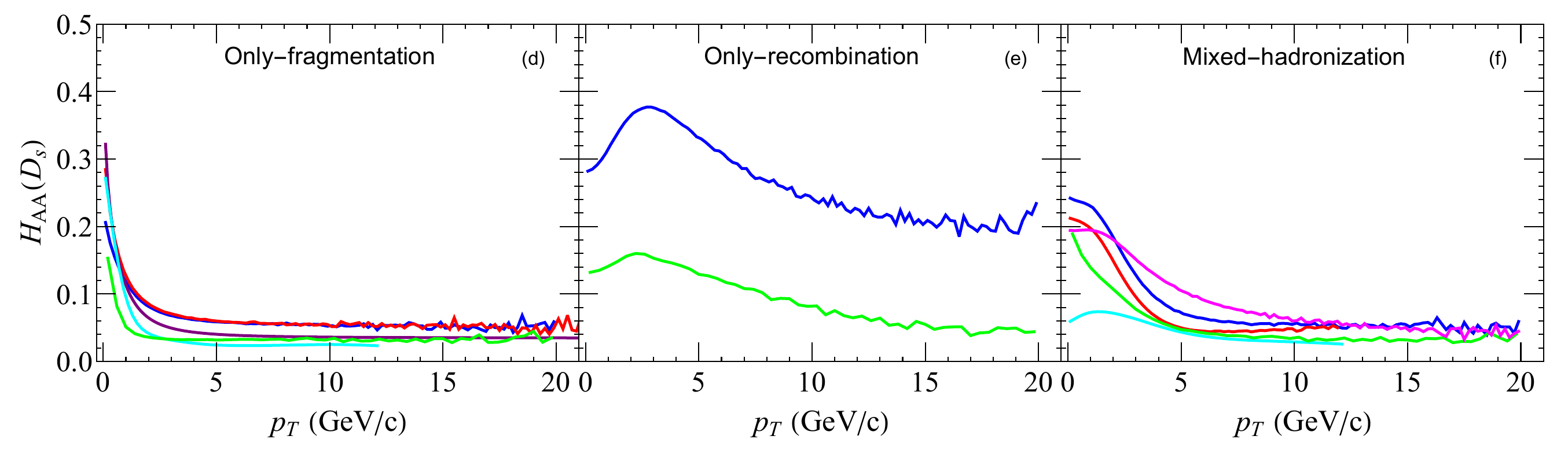}\\
\includegraphics[width=0.9\textwidth]{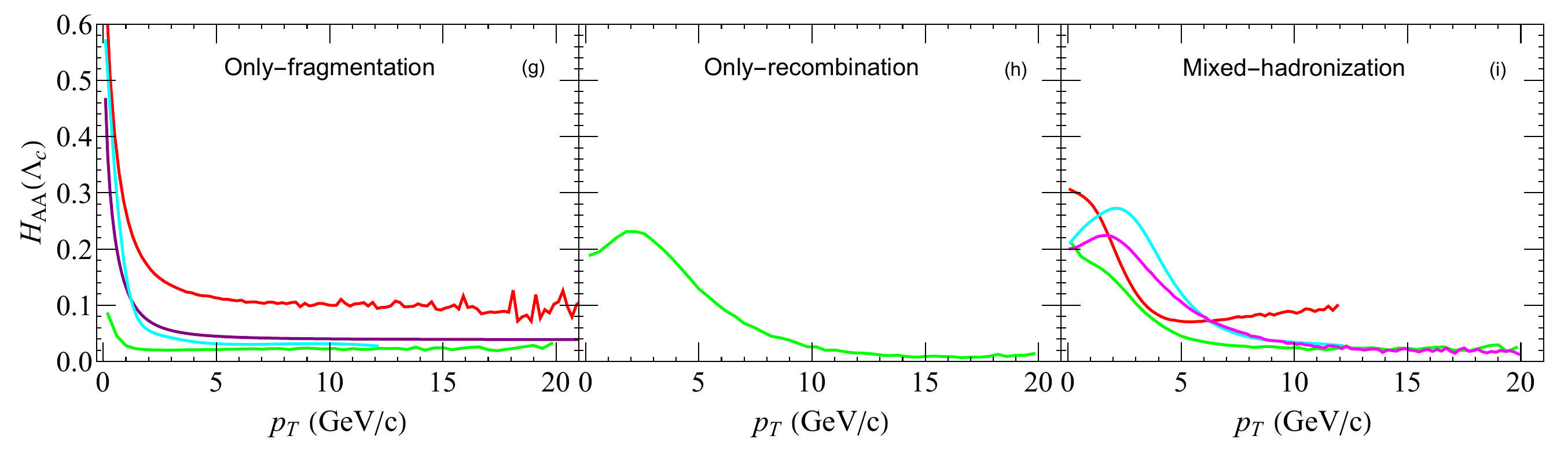}
\caption{The $H_{AA}$ of $D$, $D_s$, and $\Lambda_c$. For each charm hadron, the only-fragmentation, only-coalescence, and mixed-hadronization processes are plotted.}
\label{fig:haa}
\end{figure*}
%---------------------------------------------------------------------

%---------------------------------------------------------------------
\begin{figure*}[!htb]
\includegraphics[width=0.9\textwidth]{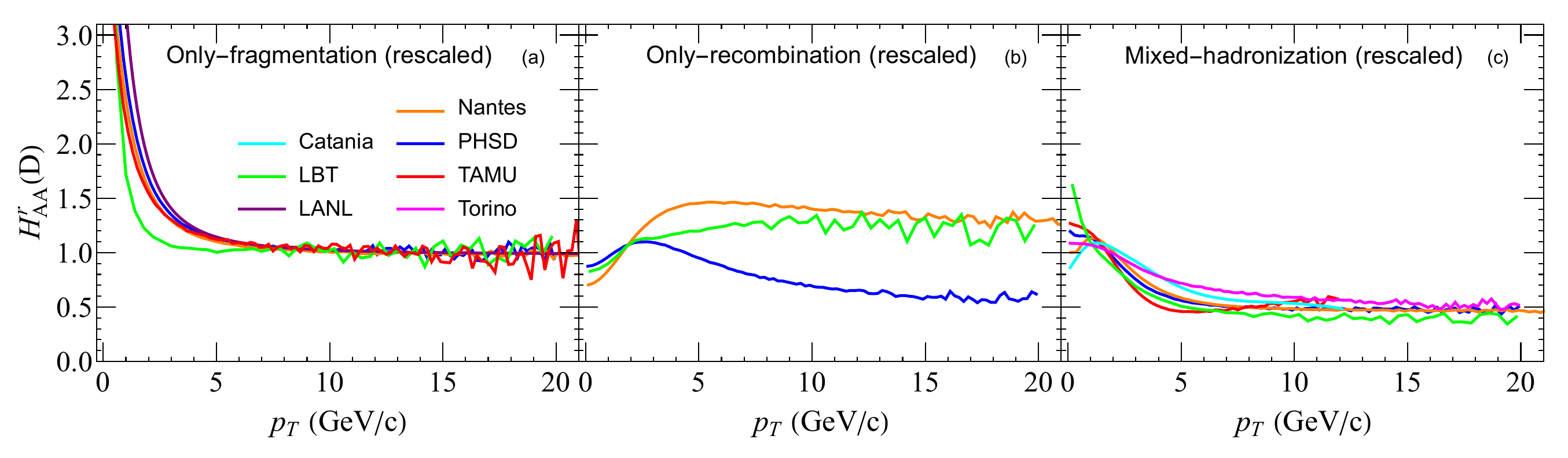}\\
\includegraphics[width=0.9\textwidth]{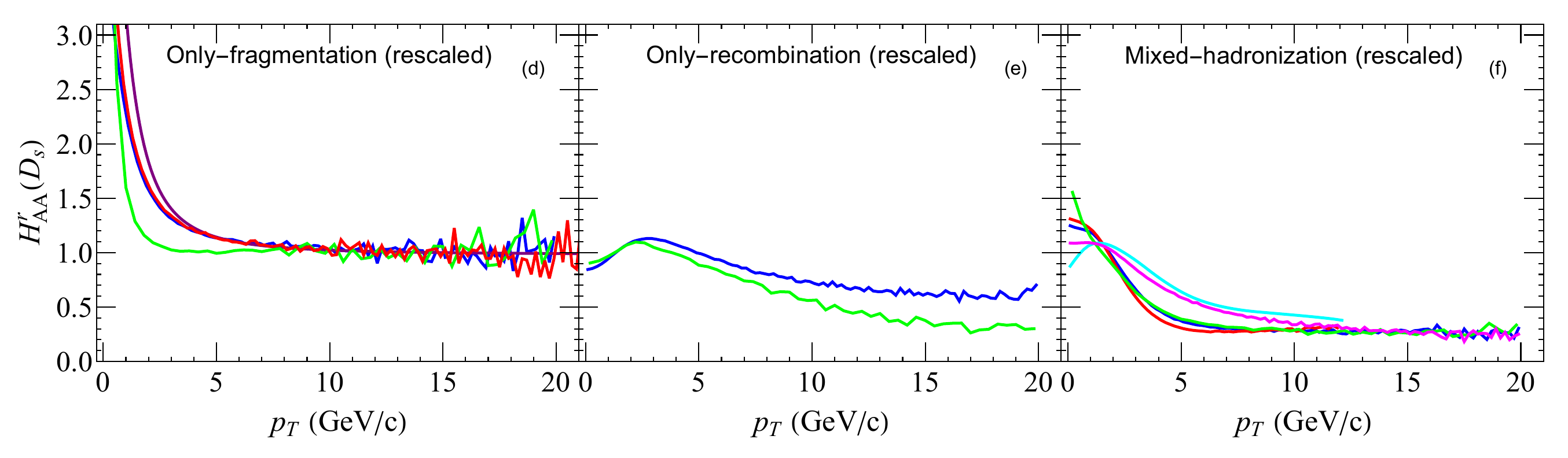}\\
\includegraphics[width=0.9\textwidth]{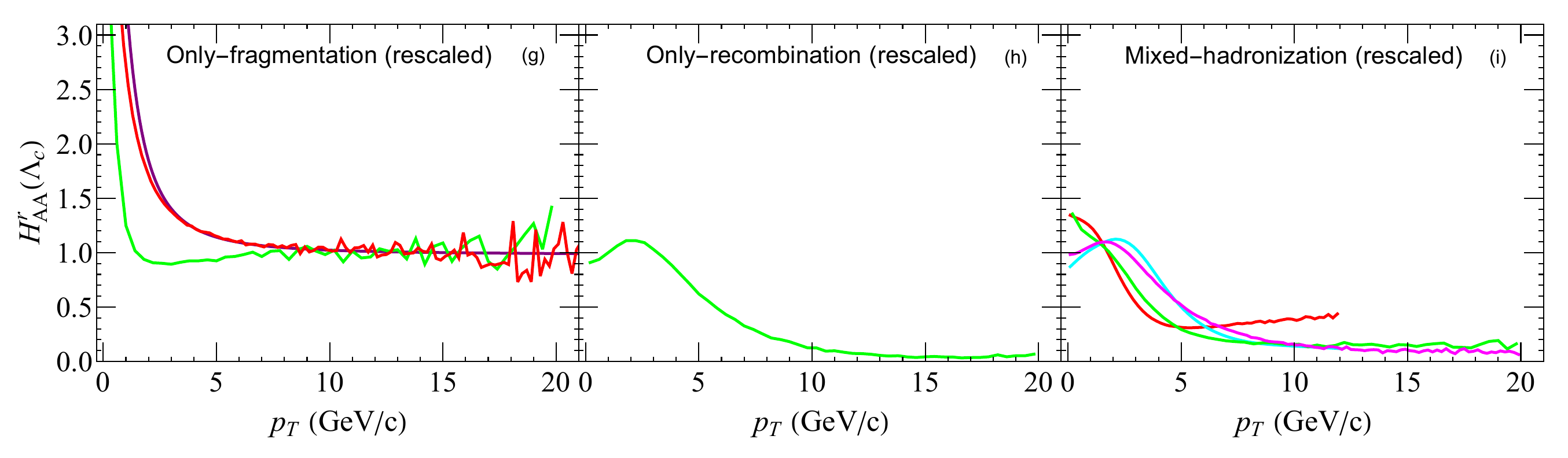}
\caption{The rescaled $H_{AA}^r$ of $D$, $D_s$, and $\Lambda_c$. For each charmed hadron, the only-fragmentation, only-coalescence, and mixed-hadronization processes are plotted.}
\label{fig3}
\end{figure*}
%---------------------------------------------------------------------

%---------------------------------------------------------------------
\begin{figure*}[!htb]
\includegraphics[width=0.9\textwidth]{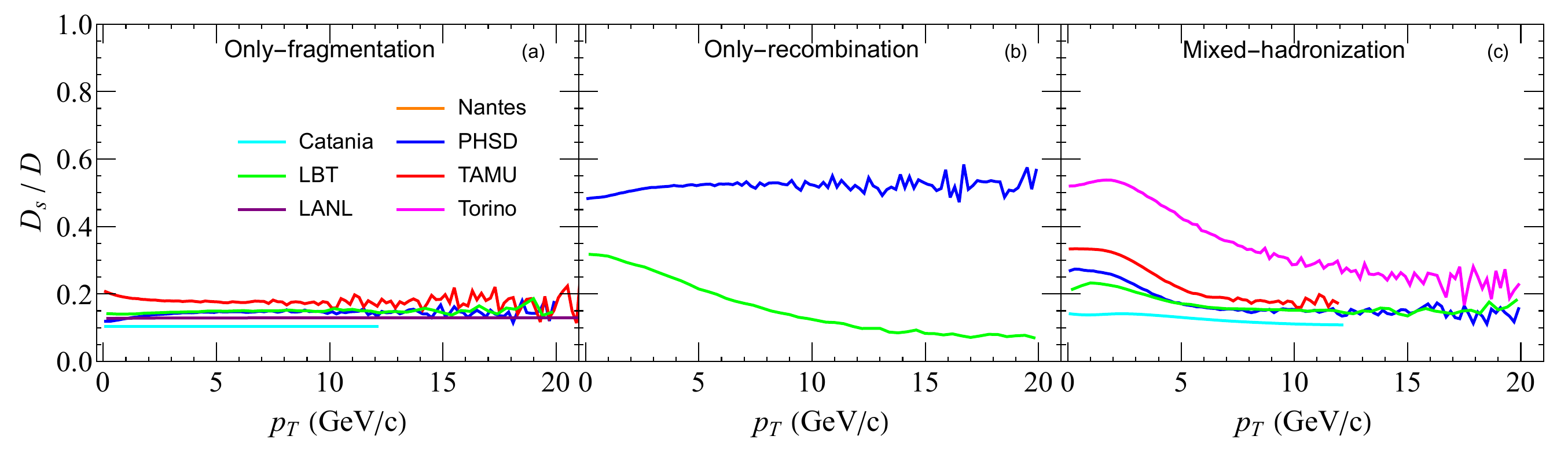}\\
\includegraphics[width=0.9\textwidth]{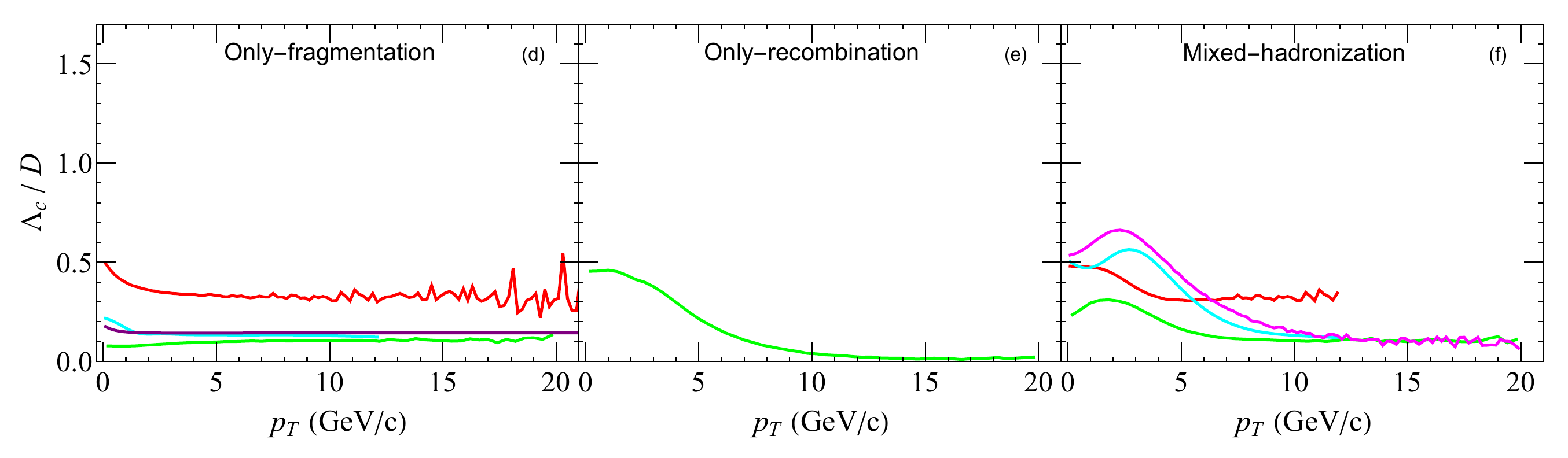}
\caption{The yield ratio of $D_s/D$ and $\Lambda_c/D$. For each charm hadron, the only-fragmentation, only-coalescence, and mixed-hadronization processes are plotted.}
\label{fig4}
\end{figure*}
%---------------------------------------------------------------------

Next, we turn to the $H_{AA}$ of $D$, $D_s$, and $\Lambda_c$ production, displayed in the top, middle, and bottom panels, respectively, of Fig.~\ref{fig:haa}. To exhibit the influence of the hadronization mechanisms we display the $H_{AA}$ for the case when all hadrons are produced exclusively via fragmentation (left), exclusively via recombination (middle), and in a combination of both (right); all results include feeddown contributions. As expected, fragmentation, where the heavy hadron has a smaller $p_T$ than the heavy parton, yields $H_{AA}$ values well below one for large $p_T$, which are directly related to the pertinent fragmentation functions, recall Fig.~\ref{fig_frag}: if the fragmentation function peaks at large values of $z$ (like that of the Nantes group), the momentum change of the heavy hadron, relative the heavy quark, is smaller and therefore the $H_{AA}$ is larger. 
The uncertainties in the fragmentation functions cause an uncertainty in the high-$p_T$ $H_{AA}$ of $D$ mesons of almost a factor of 2, and similar for the $D_s$. For the $\Lambda_c$ the models that use fragmentation functions from $e^+e^-$ collisions are in fairly good agreement, while in the TAMU model the $H_{AA}$ is larger over the entire $p_T$ range by roughly a factor of $\sim$2-3, primarily due to the large number of excited resonances which have been constrained by a fit to $pp$ data~\cite{He:2019tik}.
At low $p_T$ fragmentation functions are not reliable, and therefore the extrapolation is rather uncontrolled unless constrained by $pp$ data. Usually, this extrapolation does not affect the predicted spectra in heavy-ion collisions, since recombination dominates at low $p_T$. 

The middle panels of Fig.~\ref{fig:haa} show the results for models when enforcing hadronization by recombination only. In the Nantes model, which only considers $D$ mesons, one finds the expected coalescence effect where the addition of a light quark pushes the hadron distribution out to higher $p_T$, entailing a deficit at low $p_T$.
For a model that has more than one heavy flavor hadron, the trend of $H_{AA}$ for a specific state depends not only on the coalescence probability but also on the decay contributions. In the PHSD model resonance feeddown ``repopulates" the lower $p_T$ region at the expense of higher $p_T$ where, consequently, the $H_{AA}$ of $D$ decreases. This effect is much less pronounced in the LBT model, despite its implementation of off-shell production with subsequent pion decays.
For $D_s$, both the PHSD and LBT model show a decreasing $H_{AA}$ at large $p_T$, due to the smaller coalescence probability for $D_s$ compared to $D$~\cite{Cao:2019iqs}. 
A similar, albeit stronger, trend is observed for $\Lambda_c$ production.
However, above $p_T^c=5$ GeV, $P_{rec}$ is (well) below 0.1 for all models (except TAMU), and therefore the detailed form of $H_{AA}$ from recombination-only at high $p_T$ does not play a role for the total $H_{AA}$, which is the combination of recombination and fragmentation weighted with $P_{rec}$ and $1-P_{rec}$, respectively. 

The total (or ``mixed'') $H_{AA}$, which is a combination of the fragmentation and recombination (see Eq.~\eqref{eq.mixcoal}), is shown in the right panel of Fig.~\ref{fig:haa}.
Its high-$p_T$ behavior is almost identical to the fragmentation-only $H_{AA}$. However, towards lower $p_T$ the increase of $H_{AA}$ is now due to the recombination contribution.  At still lower $p_T$, most models develop a leveling-off in the total $H_{AA}$,
or even a slight maximum structure in the Nantes model as a consequence of the stronger effect in the recombination part (see middle panel).
In the TAMU, LBT, and PHSD models, a ``flow bump" is not generated for the given fireball configuration, pointing at differences in how the parton collectivity (both the charm quark and the collective flow of the medium) is converted into the one of hadrons.

Even if in some models
a heavy quark with $p_T^c=0$ hadronizes exclusively via recombination, $D$-mesons with $p_T^D=0$ can also originate from fragmentation of $c$-quarks with a finite $p_T^c$. Therefore, at $p_T^D=0$ we see a difference between the pure recombination and the mixed scenario.    
For $D_s$ mesons, a fair agreement of the $H_{AA}$'s is found, except for the Catania model which falls short of the other models by a factor of 2-3, as was borne out already of the inclusive fractions listed in Tab.~\ref{tab_frac}. 
For $\Lambda_c$ baryons, one finds an appreciable sensitivity to the recombination kinematics at low $p_T$, even though the inclusive fractions are not very different (between 15\% and 24\%, cf.~Tab.~\ref{tab_frac}). At higher $p_T$ the TAMU model sticks out due to the large feeddown contribution from excited baryons.

The model differences in the charm hadro-chemistry are a substantial part of the discrepancies found in Fig.~\ref{fig:haa}. In an attempt to eliminate this effect and focus on the kinematics, we display in Fig.~\ref{fig3} a rescaled ratio, 
$H_{AA}^r \equiv H_{AA}/R$ with
\begin{eqnarray}
R={\int (dN_c/dp_T)H_{AA}dp_T \over \int (dN_c/dp_T) dp_T}
\end{eqnarray}
for coalescence-only and mixed-hadronization processes, and 
\begin{eqnarray}
R={\int_{10}^{20} (dN_c/dp_T)H_{AA}dp_T \over \int_{10}^{20} (dN_c/dp_T) dp_T}
\end{eqnarray}
for fragmentation-only processes, where the integration limits are chosen under the premise that at high $p_T$ fragmentation functions can be reasonably well adjusted to experimental data.  We observe that the fragmentation-only $H_{AA}^r$'s (left panels) at high $p_T$ now rather closely agree for $p_T\gtrsim8$\,GeV among all models for all 3 hadron types. Even for at low $p_T\lesssim5$\,GeV, the agreement is not bad (with an outlier from LBT), which, however, may not be very significant in practice. 
For the recombination-only scenario (middle panels in Fig.~\ref{fig3}), the model curves are now closer together, but the features in the $p_T$ dependence remain as discussed in the context of Fig.~\ref{fig:haa}.

For the ``mixed'' $H_{AA}^r$, the $D$-meson results are now rather similar (in contrast to the $H_{AA}$ in Fig.~\ref{fig:haa}), with values around 1 at low $p_T$, where most of the normalization is operative. 
The shape differences are more pronounced for $D_s$ and $\Lambda_c$, where two classes seem to emerge, i.e., Catania and Torino vs. PHSD, TAMU and LBT.
Somewhat surprisingly, the high-$p_T$ $H_{AA}^r$ for the mixed case shows a similar convergence between the models for all hadron species (with the exception of the $\Lambda_c$ from TAMU) as seen in the fragmentation only case (left column), despite a rather different normalization.  

Next, we study the recombination and fragmentation probabilities for different charmed hadrons, by plotting in Fig.~\ref{fig4} the yield ratios $D_s/D$  (top row) and $\Lambda_c/D$ (bottom row) as a function of $p_T$. Here $D$ means $D^0$ plus $D^+$.  For fragmentation-only (left panels) we see an almost constant ratio for all approaches, indicating that the form of the fragmentation functions are very similar. The absolute values differ, however, slightly for $D_s$, while for $\Lambda_c$ the TAMU model gives a twice larger multiplicity than in all other models (again due to charm-baryon resonance feed-down adjusted to reproduce to $pp$ data).
For the recombination-only scenario, we only have two results for $D_s$, which differ quite a bit. The LBT results show an increased ratio towards low $p_T$, presumably a consequence again of the pion decay feeddown, whereas for PHSD a falling trend is found characteristic of recombination with little or no feeddown.
For the mixed scenario (fragmentation plus recombination), with the weights plotted in Fig.~\ref{fig1}, all $D_s/D$ results show an increase over the only-fragmentation scenario, especially at low $p_T$, and in all approaches this is likely due to the enhanced $s$-quark content in the QGP, most pronounced for the local color neutralization mechanism (Torino), followed by TAMU (including resonance feeddown), and less for Catania, LBT, and PHSD which use Wigner functions. For the $\Lambda_c/D$ ratio, the  Torino and TAMU results are rather large again, while for Catania the tuning of the fragmentation function to $pp$ data also translates into a large value in the heavy-ion environment, while the LBT result, with a fragmentation function tuned to $e^+e^-$ collisions, remains rather small, similar to the $e^+e^-$ results. The TAMU result becomes the largest at high $p_T$  due to the feeddown of heavy charm-baryon resonances present in their fragmentation function.

%---------------------------------------------------------------------
\begin{figure*}[!htb]
\includegraphics[width=0.6\textwidth]{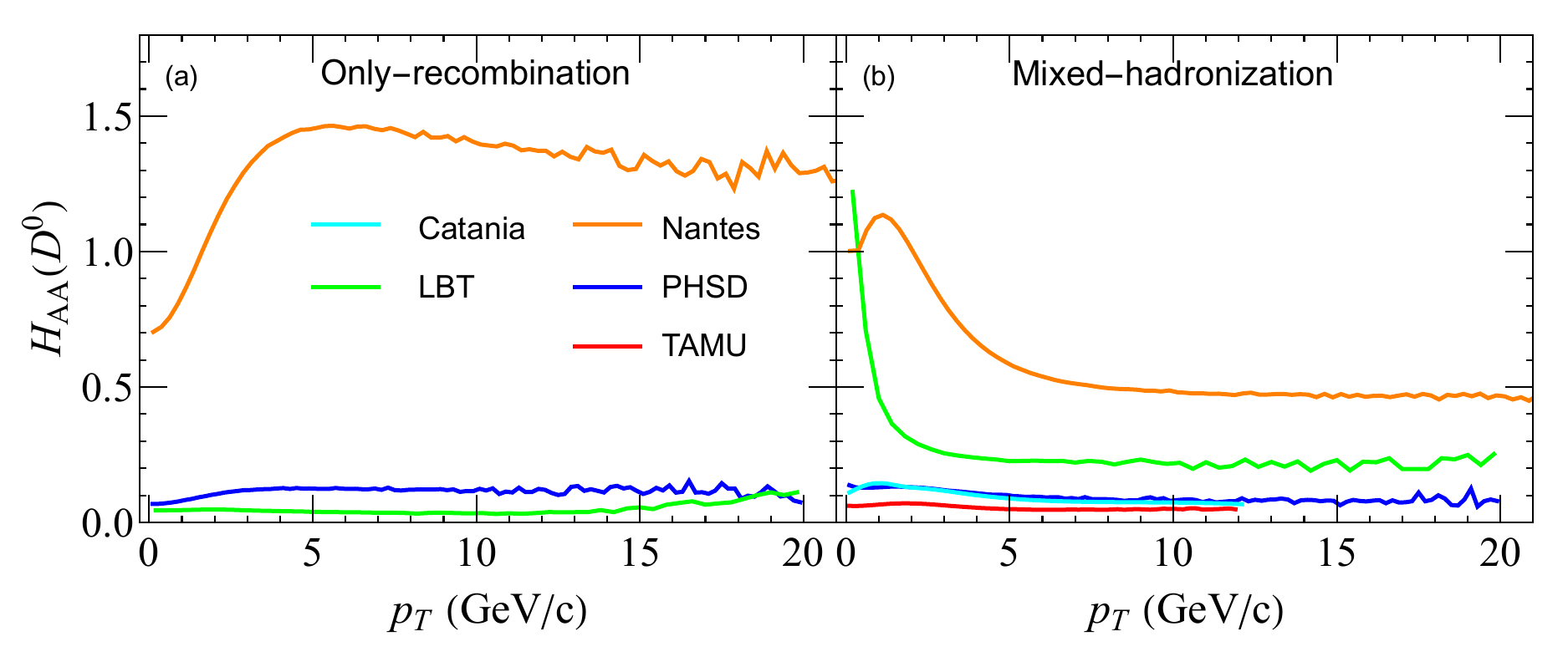}\\
\includegraphics[width=0.6\textwidth]{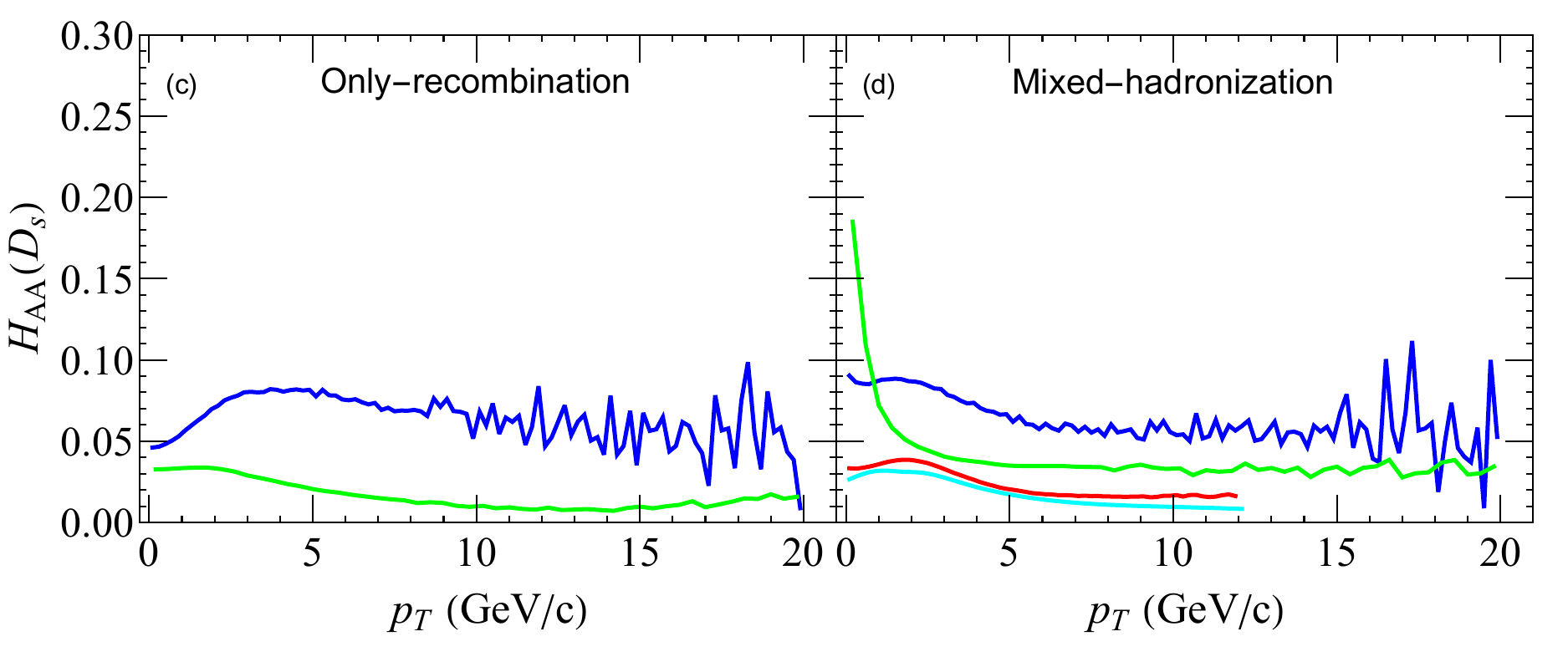}\\
\includegraphics[width=0.6\textwidth]{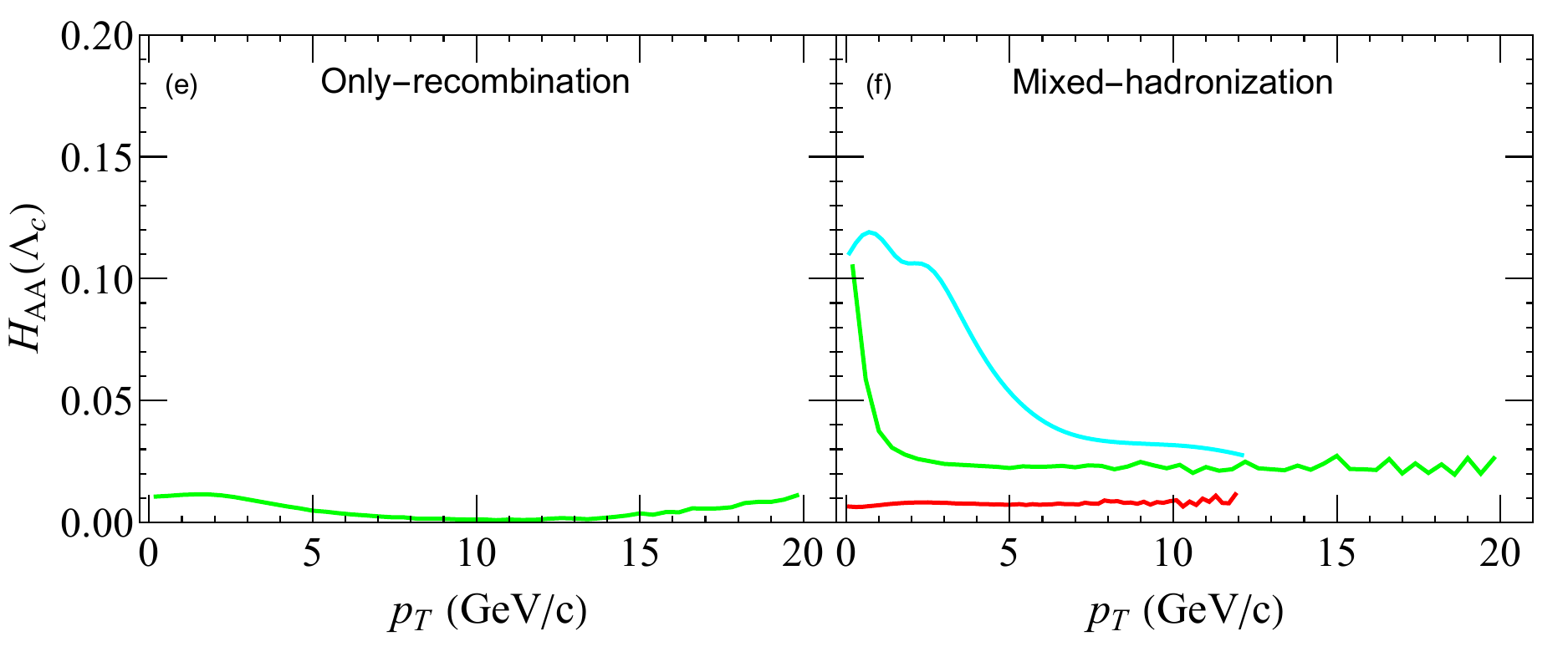}
\caption{The $H_{AA}$ of the direct $D^0$, $D_s$, and $\Lambda_c$. For each charmed hadron, the only-coalescence and mixed-hadronization processes are plotted.}
\label{fig7}
\end{figure*}
%---------------------------------------------------------------------

To eliminate the feeddown from excited states, which vary largely throughout the models,
we compare in Fig.~\ref{fig7} the $H_{AA}$'s for directly produced hadrons, to be confronted with the inclusive results in Fig.~\ref{fig:haa}. 
The left column shows the results for recombination only whereas the right column contains the combined results for recombination plus fragmentation. From top to bottom, we plot the results for $D$, $D_s$, and $\Lambda_c$.  
For the two recombination-only results, LBT and PHSD, the $H_{AA}$ value for both $D^0$ and $D_s$ drops markedly compared to the prompt ones plotted in Fig.~\ref{fig:haa}. This points to the importance of the excited states in both models. The decay of excited states shifts the heavy-meson momentum. From the near-constant $H_{AA}$ as a function of $p_T$ for directly produced heavy hadrons observed in PHSD, one may postulate that the decrease of  $H_{AA}$ towards large $p_T$ observed for prompt hadrons is due to resonance decay. However, such an effect is not observed in the LBT approach, calling for a more detailed investigation of the role of excited states. 
For the mixed-hadronization, a similar pattern is observed, but also here the degree of feed-down varies substantially so that the previously decent agreement for the prompt $H_{AA}$'s fans out substantially (where TAMU drops the most, as expected due to its larger extensive feed-down). 
Thus, settling the issue of 
feeddown contributions will play a critical role in establishing better consensus in the model approaches. 

%---------------------------------------------------------------------
\begin{figure*}[!htb]
\includegraphics[width=0.6\textwidth]{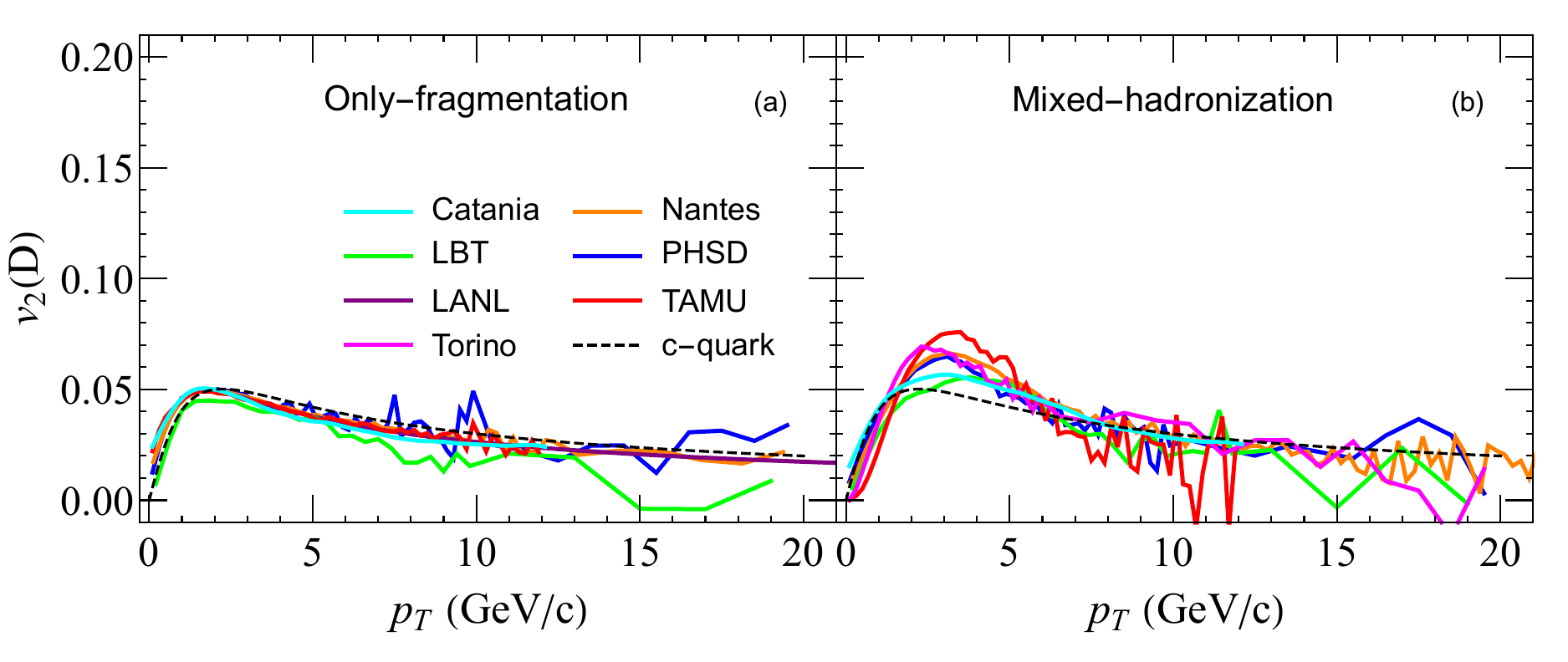}\\
\includegraphics[width=0.6\textwidth]{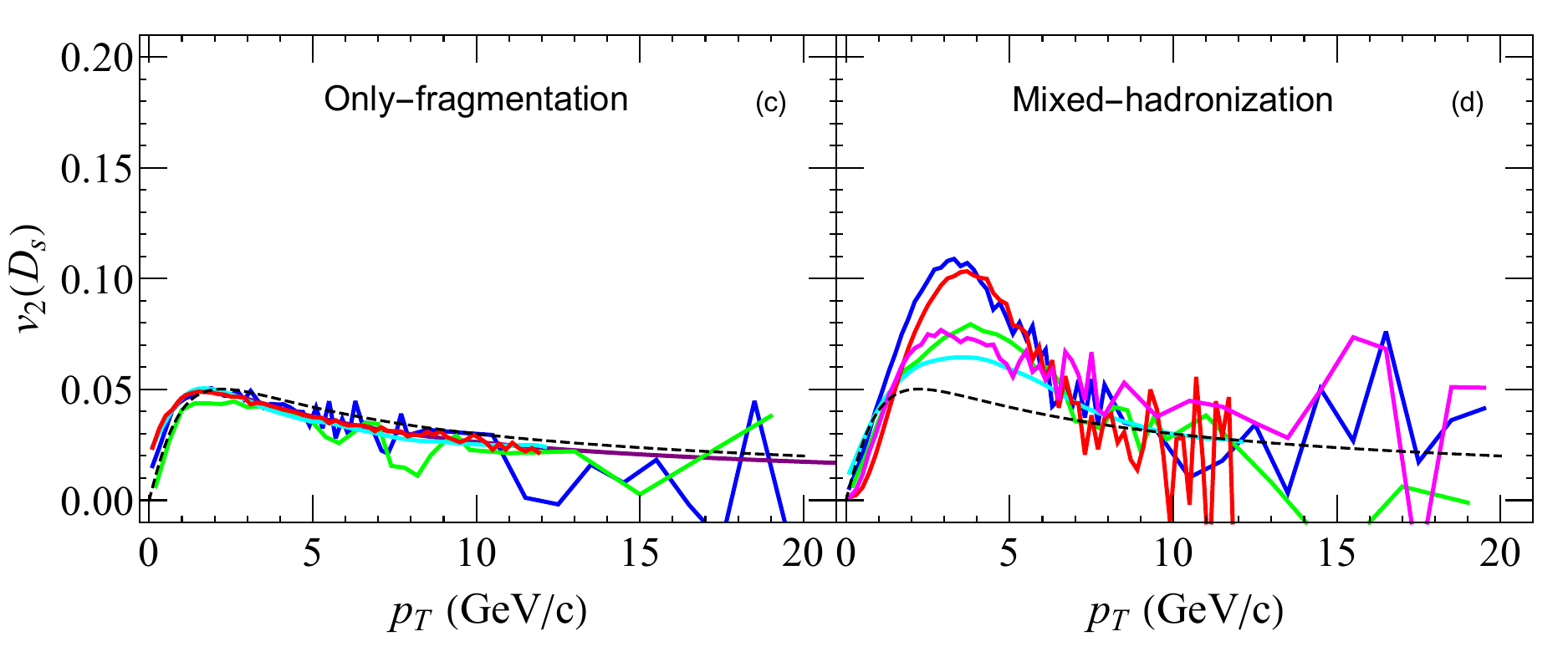}\\
\includegraphics[width=0.6\textwidth]{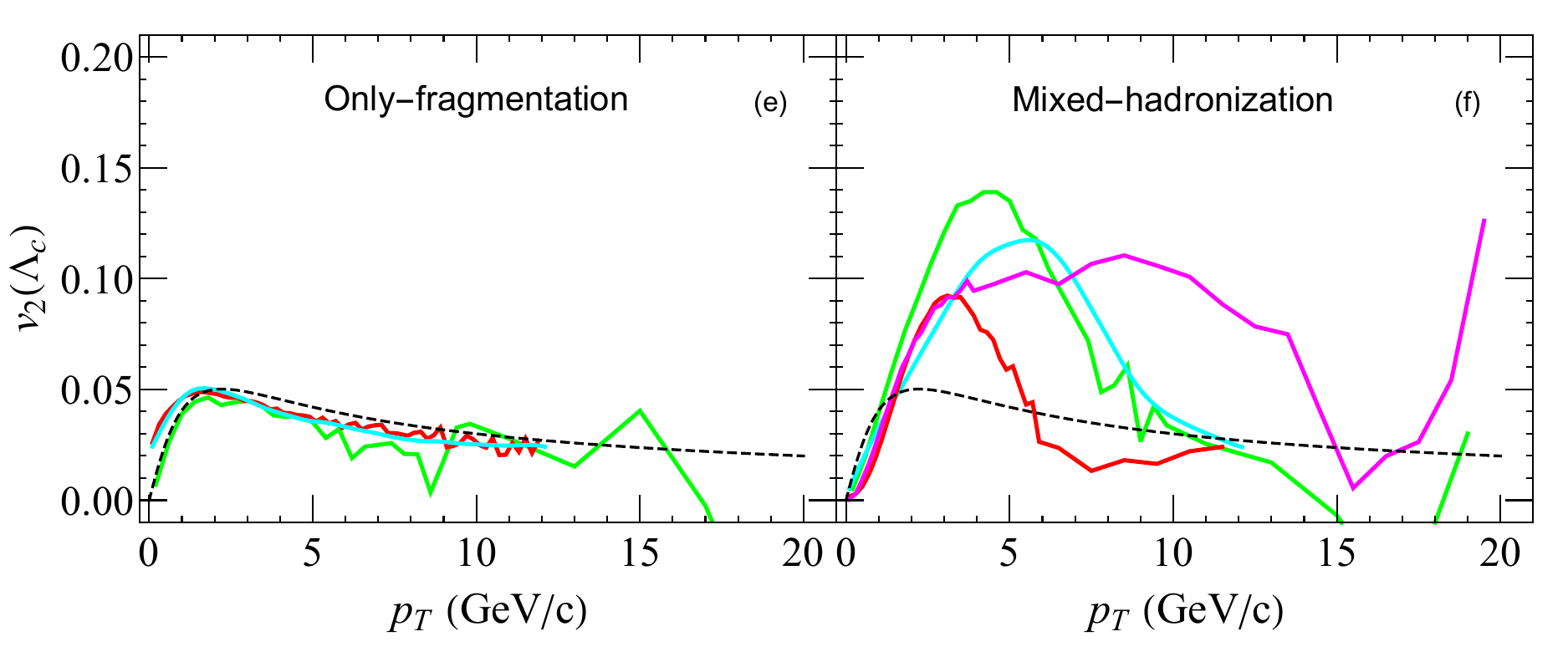}
\caption{The elliptic flow, $v_2$, of $D$, $D_s$, and $\Lambda_c$  (including feeddown contributions) for a $c$-quark distribution given by Eq.~\eqref{eq:v2c}. For each charm hadron, the only-fragmentation (left panels) and mixed-hadronization (right panels) scenarios are displayed.}
\label{fig:v2}
\end{figure*}
%---------------------------------------------------------------------

%%%%%%%%%%%%%%%%%%%%%%%%%%%%%%%%%%%
\subsection{Elliptic Flow of Charm Hadrons}
\label{ssec_hadron-v2}
%%%%%%%%%%%%%%%%%%%%%%%%%%%%%%%%%%
Another quantity, which is critical in the interpretation of experimental results, is the elliptic flow, $v_2$, the second-order coefficient of the Fourier analysis of the azimuthal distribution of produced particles. Heavy quarks, initially produced in a hard process and therefore commonly assumed to carry vanishing $v_2$ (although that may not necessarily be correct in the presence of initial state correlations), are expected to acquire a finite $v_2$ by interactions with light partons from the medium, as the latter develop a finite $v_2$ during the expansion of the QGP due to an initial spatial eccentricity of the interaction zone. In addition, when the heavy quark recombines with a light quark to form a heavy meson, the underlying kinematics implies that the heavy mesons inherit further $v_2$ contributions from the light quarks (strictly speaking, these 2 contributions are not necessarily independent~\cite{He:2022ywp}). It is therefore desirable to assess these two contributions individually and to understand how the hadronization process modifies the  $v_2$.  Here we focus on how the $v_2$ of the heavy meson is generated during hadronization. However, since no straightforward decomposition such as Eq.~\eqref{eq.mixcoal} can be given for the $v_2$, considering the $v_2$ associated with an only-coalescence setting turns out to be problematic for the present analysis. We therefore focus on the well-defined cases of only-fragmentation and mixed hadronization.

We start by displaying in Fig.~\ref{fig:v2} the $v_2$ as a function of $p_T$
for $D^0$ (top row), $D_s$ (middle row), and $\Lambda_c$ (bottom row) including feeddown contributions, for the only-fragmentation scenario (left column) and the mixed-hadronization (right column). 
The color coding is the same as in Fig.~\ref{fig_frag}, supplemented with a dashed line to benchmark the $v_2$ of the input charm-quark distribution. In almost all models, the only-fragmentation $v_2$ of the heavy hadrons is very similar to that of the heavy quark. This is expected because the fragmentation into hadrons is collinear with the HQ direction. 
In the mixed scenario, we observe $v_2(\Lambda_c)>v_2(D_s)>v_2(D)$ for basically all models. This hierarchy is mainly caused by two reasons: (a) The $v_2$ of $\Lambda_c$ is larger than that of $D_s$ and $D$ due to an additional light quark picked up via recombination; (b) The $v_2$ of $D_s$ is larger than that of $D$ due to a smaller fraction of $D_s$ produced by fragmentation (relative to $D^0$), or, in other words, due to the larger recombination fraction facilitated by a higher concentration of strange quarks in the QGP relative to $pp$ collisions~\cite{He:2012df}. Passing from $D \to D_s \to \Lambda_c$, the agreement between the models becomes worse.
For the $\Lambda_c$, the extent to which the $v_2$ is increased over the meson-$v_2$ is close to a factor of 2 for Catania, LBT, and Torino, while it is less for TAMU due to its relatively large fragmentation fraction, recall Fig.~\ref{fig:haa}. Thus, the relative increase of the baryon $v_2$ over the meson $v_2$ could potentially provide a measure of the fragmentation fraction of charm baryons in Pb-Pb collisions.

\begin{figure*}[!htb]
\includegraphics[width=0.9\textwidth]{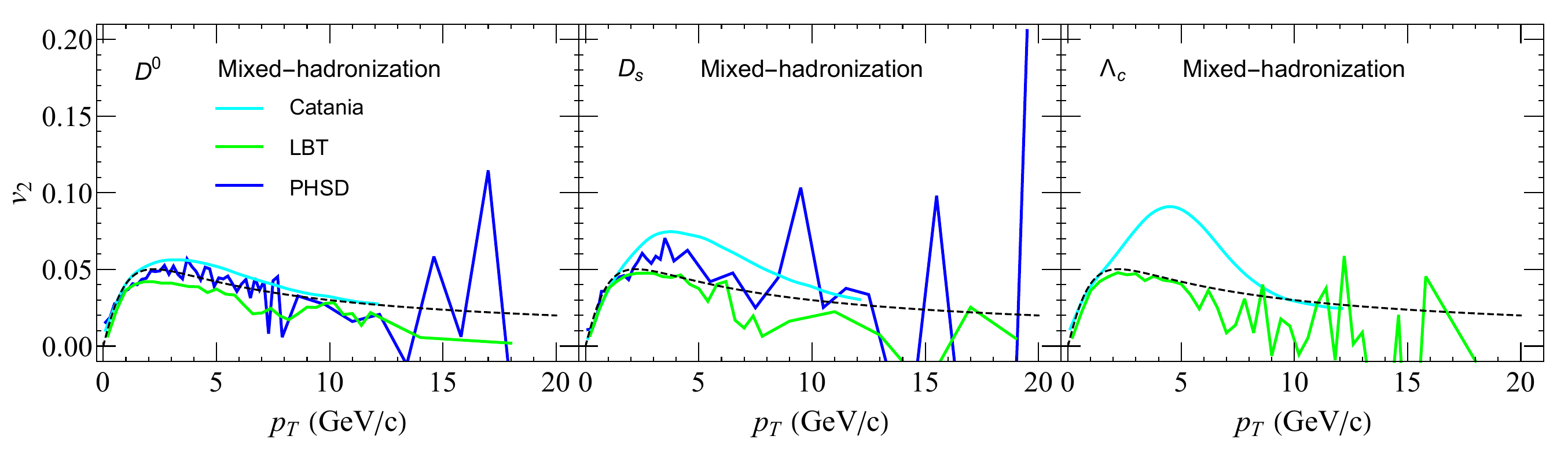}
\caption{The elliptic flow, $v_2$, of directly produced $D^0$ (left panel), $D_s$ (right panel), and $\Lambda_c$ (right panel) using the $c$-quark distributions defined in Eq.~\eqref{eq:v2c}. For each charmed hadron, the results of the mixed hadronization process are plotted.}
\label{fig8}
\end{figure*}

Finally, to have a more direct comparison of basic kinematics in $v_2$ production within the various models, we eliminate the effect of feeddown contributions and collect the results for the $v_2$ of the directly produced hadrons in Fig.~\ref{fig8}.  
First, we recall that $v_2$ from the fragmentation process is essentially the same as that of the charm quark. Compared to Fig.~\ref{fig:v2}, we see that directly produced hadrons have a considerably lower $v_2$ than for the case including feeddown from excited states. 

%---------------------------------------------------------------------
%---------------------------------------------------------------------

In the real transport calculations that include the full QGP evolution, strong momentum-coordinate space correlations of the heavy quarks with the surrounding medium should be taken into account~\cite{He:2019vgs,Beraudo:2022dpz}. 
As model studies show, these correlations may enhance $v_2^D$ of the order of 50\% in certain $p_T$ regions, especially at intermediate $p_T$ (more precisely in the transition region from recombination to fragmentation processes) where the mere increase of the recombination yield is the leading effect~\cite{He:2019vgs,Beraudo:2022dpz}.

%==================================
\section{Conclusions}
\label{sec_concl}
%==================================
In this paper, we have conducted specific comparisons between existing models of the hadronization of charm quarks. Hadronization is required to connect the computations of their transport through the QGP to experimental observables.
Specifically, we have investigated how the yields and momentum spectra of produced HF hadrons emerge from the same underlying charm-quark distribution that includes both radial and elliptic flow. Most models employ a combination of recombination and fragmentation mechanisms, but the concrete implementations differ substantially. From the comparisons of the resulting hadron spectra, the following observations can be made:

\begin{itemize}
\item 
The recombination probability, $P_{rec}(p^c_T)$,  and hence the fragmentation probability $1-P_{rec}(p^c_T)$, varies strongly in the different approaches. An important cause of this difference is likely the role of excited states. In some models only ground-state hadrons are included, while others include a rich spectrum of excited states (in particular for baryons), predicted by relativistic quark models. This also extends the $p_T^c$ reach of recombination substantially, which is especially important for the intermediate $p_T^c$ region where recombination and fragmentation mechanisms compete. Finally, different normalization conditions are applied for the low-momentum limit of recombination. 
\item 
The shift in $p_T$ during the hadronization process, defined as $p_T^D-p_T^c$, is generally positive for recombination and negative for fragmentation processes. Surprisingly, the model variation in the former turns out to be smaller than in the latter (recall the comparison of Fig.~\ref{fig_frag} vs.~Fig.~\ref{fig_shift} (left panel)).
\item 
The functional form of the hadronization ratio, $H_{AA}(i)={dN_i/dp_T \over dN_c/dp_T}$, at high $p_T$ is mostly controlled by the excited states which are included in the fragmentation process, but exhibits the largest model variations at small $p_T$, especially when split up into different hadron species. Also here the excited states play a key factor for the difference, together with the model assumptions for $P_{rec}(p_T^c)$ in the small-momentum limit. The difference between the various recombination models in how the collectivity of the $c$-quarks is converted into that of charm hadrons
becomes particularly apparent when focusing on direct production (i.e., no feeddown), where different shapes of the $H_{AA}$ at low $p_T$ exhibit the varying kinematics in the recombination models.
\item 
The elliptic flow of the fragmentation component of charm hadrons agrees fairly well among the models, closely coinciding with that of the parent $c$-quark. The elliptic flow from the full hadronization models also agrees on a qualitative mass hierarchy for the maximal $v_2$ of $D$, $D_s$, and $\Lambda_c$, although the relative enhancement differs somewhat, especially for the $\Lambda_c$, with the fragmentation fraction (and its feeddown contributions) playing an important role. For this observable the decay feeddown also appears to play a crucial role.
\end{itemize} 

Future progress on these issues will likely require a more systematic resorting to basic principles (e.g., by developing closer  connections to the properties of the QGP) and constraints, as well as the identification of suitable benchmarks in model comparisons (e.g., by scrutinizing the direct production of charm hadrons). 
Constraints from $pp$ collisions, including multiplicity dependencies, for which high-precision data will become increasingly abundant, may help disentangle contributions from fragmentation and recombination processes. This could be very effective to constrain the hadro-chemistry, both inclusive and as a function of $p_T$ and thus to reduce uncertainties associated with feeddown. The latter turned out to be a major source of the model variations in hadron observables, largely governed by the $p_T$-dependent recombination fraction which controls the interplay of recombination and fragmentation. More ambitious constraints can be envisaged from lattice QCD, such as in-medium effects on the partonic and hadronic properties in the recombination process. 

\vspace{1cm}
\noindent {\bf Acknowledgement}: 
We acknowledge support by the Deutsche Forschungsgemeinschaft (DFG, German Research Foundation)
through the grant CRC-TR 211 ’Strong-interaction matter under extreme conditions’ - Project number
315477589 - TRR 211. This work is supported by the European Union’s Horizon 2020 research and innovation
program under grant agreement No 824093 (STRONG-2020), and 
by the U.S. Department of Energy, Office of Science, Office of
Nuclear Physics through the Topical Collaboration in Nuclear Theory on Heavy-Flavor Theory (HEFTY) for QCD Matter under award no. DE-SC0023547.
RR has been supported by the US National Science Foundation under grant nos. PHY-1913286 and PHY-2209335.
SAB and WF acknowledge support by the National Science Foundation (NSF) within the framework of the JETSCAPE collaboration, under grant number OAC-2004571 (CSSI:X-SCAPE) and ACI-1550225 as well as by the Department of Energy under  grant number DE-FG02-05ER41367.
VG, VM and SP acknowledge the support of linea 2 UniCT Grant for the HQCDyn project. SC is supported by the National Natural Science Foundation of China (NSFC) under Grant Nos. 12175122 and 2021-867. IV is supported in part by the Laboratory Directed Research and Development  (LDRD) program at LANL.

%\clearpage
%=====================
\bibliographystyle{apsrev4-1.bst}
\bibliography{Ref}

\end{document}